\documentclass[a4paper,fleqn]{aa}

\pdfoutput=1

\usepackage{natbib}
\usepackage{txfonts}
\usepackage{gensymb}
\usepackage{inputenc}
\usepackage{float}
\usepackage{multicol}
\usepackage{multirow}

\bibpunct{(}{)}{;}{a}{}{,}

\usepackage{color}
\usepackage{url}
\usepackage[dvipsnames]{xcolor}

\usepackage{caption}
\usepackage{subcaption}

\usepackage{ulem}

\usepackage[breaklinks=true,colorlinks,citecolor=blue]{hyperref}

\usepackage{breakurl}
\usepackage{graphicx}
\defcitealias{Siudek2018a}{S1}

\usepackage{soul}
\usepackage{verbatim}	

\begin{document}

\title{
   Shaping physical properties of galaxy subtypes in the VIPERS survey: environment matters}
\titlerunning{Environment within galaxy subclasses}

\author{M. Siudek\inst{1,2}\thanks{E-mail: msiudek@ifae.es}, K. Małek\inst{2,3}, A. Pollo\inst{2,4}, A. Iovino\inst{5}, C. P. Haines\inst{6}, M.~Bolzonella\inst{7}, O.~Cucciati\inst{7}, A.~Gargiulo\inst{8}, B.~Granett\inst{5}, J.~Krywult\inst{9}, T.~Moutard\inst{3,11}, M.~Scodeggio\inst{10}}
\institute{
$^{1}$Institut de F’isica d’Altes Energies, The Barcelona Institute of Science and Technology, 08193 Bellaterra, Spain\\
$^{2}$National Centre for Nuclear Research, ul. Pasteura 7, 02-093 Warsaw, Poland\\
$^{3}$Aix Marseille Univ, CNRS, CNES, LAM, Marseille, France\\
$^{4}$Astronomical Observatory of the Jagiellonian University, ul. Orla 171, 30-244 Krak\'ow, Poland\\
$^{5}$INAF - Osservatorio Astronomico di Brera, Via Brera 28, 20122 Milano --  via E. Bianchi 46, 23807 Merate, Italy \\
$^{6}$Instituto de Astronom\'{i}a y Ciencias Planetarias de Atacama (INCT), Universidad de Atacama, Copayapu 485, Copiap\'{o}, Chile\\
$^{7}$INAF - Osservatorio di Astrofisica e Scienza dello Spazio di Bologna,via Gobetti 93/3, 40129 Bologna, Italy\\
$^{8}$INAF - Istituto di Astrofisica Spaziale e Fisica Cosmica Milano, via A.Corti 12, 20133 Milano, Italy\\
$^{9}$Institute of Physics, Jan Kochanowski University, ul. Swietokrzyska 15, 25-406, Kielce, Poland\\
$^{10}$INAF - Istituto di Astrofisica Spaziale e Fisica Cosmica Milano, via Bassini 15, 20133 Milano, Italy\\
$^{11}$Department of Astronomy \& Physics and the Institute for Computational Astrophysics, Saint Mary’s University, 923 Robie Street, Halifax, Nova Scotia, B3H 3C3, Canada\\}
\authorrunning{M. Siudek et al.}

	%
  \abstract
{}
{This study aims to explore the relation between the physical properties of different galaxy subclasses, from red passive to blue star-forming, and their environment. Our work is based on the analysis of 31\,631 VIMOS Public Extragalactic Redshift Survey (VIPERS) galaxies observed at $0.5<z<0.9$.  
The unprecedented volume of VIPERS and the wealth of auxiliary derived data allow us to associate subclasses of the main galaxy populations with their possibly different evolutionary paths. Such a study is performed for the first time with such statistical precision. 
 }
{We use the results of an unsupervised clustering algorithm  to distinguish 11 subclasses of VIPERS galaxies based on the multi-dimensional feature space defined by rest-frame $UV$ to $NIR$ colours presented in~\cite{Siudek2018a}.  
We investigate the relationship between the properties of these subclasses of galaxies  and their local environment, defined as the galaxy density contrast, $\delta$, derived from the $5^{th}$ nearest neighbour technique.  
}
{
We confirm  that the galaxy population-density relation is already in place at $z\sim0.9$, with the blue galaxy fraction decreasing with density, compensated by an increase of the  red fraction. 
We demonstrate how the properties of red, green, and blue galaxy subclasses are altered as they assemble into denser regions and we attempt to interpret it in the context of their evolution. 
On average red galaxies in the high-density environment are larger by $28\%$ than the ones in low-density environments. 
In particular, we find one group of galaxies, subclass C3, whose increase of size with time can be explained mainly as the result of mergers; for other red subclasses, mergers would not seem to play the major role (subclass C2) or play a negligible role (subclass C1). 
The properties of the green galaxies (subclasses C4--6) depend on whether their stellar mass is above or below a transition mass, $\log ({\rm M}_{\rm star}/{\rm M}_{\odot})=10.6$. 
Low-mass green ($9.5\lesssim\log ({\rm M_{star}/M}_{\odot})\lesssim10.6$) galaxies appear to have grown through secular processes, while in high-mass ($10.6\lesssim\log ({\rm M_{star}/M}_{\odot})\lesssim11.5$) green galaxies mass assembly appears to be dominated by mergers. 
When it comes to blue galaxies, the trend of decreasing fraction with denser environments seen for the group as a whole (subclasses C7--11) is found to be driven mostly by one (the most numerous) group of galaxies, subclass C10. 
These are compact low-mass galaxies with high specific star formation rates, that are preferentially found in low-density environments. 
However, the remaining blue galaxies (subclasses C7--9) are larger and appear in denser environments than galaxies within C10.  
}
{}


\maketitle


	\section{Introduction}\label{sec:introduction}

In the paradigm of hierarchical structure formation, the evolution of the primordial density field acting under gravitational instability drives dark matter to cluster and collapse into virialised objects (haloes). 
Such haloes provide the potential wells into which baryons fall and galaxies subsequently form~\citep{white1978}.  
Since we observe different populations of galaxies (spiral, elliptical, irregular) and we know that their distribution in the large scale structure is not random, we can expect the local galaxy density field and/or properties of the host dark matter halo have to influence the evolution of galaxy properties. 
There are two main possible external factors affecting the evolution of galaxies in the large-scale structure of the Universe: properties of the host dark matter halo (its mass and angular momentum) and interactions with other galaxies. 
The importance of these two factors is often discussed in the context of the so-called nature vs nurture scenario, where nature corresponds to host dark matter halo properties~\citep{Eggen1962} and nurture - to the interactions with other surrounding galaxies~\citep[e.g.][]{Toomre1972}. 
Presently, it seems to be already clear that the galaxy evolution was driven by a mixture of both factors~\citep[e.g.][]{peng2010}. 
However, their role in the evolution of particular populations of galaxies and timescales of the corresponding processes are not yet clear.

Studying relations between morphology and environment of galaxies in the local Universe \cite{Dressler1980} found that most elliptical galaxies are located in the densest  environments, such as groups and clusters, whereas most spiral galaxies are found in less dense areas. 
The morphological segregation of galaxies with density appears to be a universal characteristic of galaxy populations~\citep[e.g.][]{Balogh1997, Lewis2002,Hogg2003,Kauffmann2004}. 
Such segregation was confirmed later also beyond the local Universe~\citep[e.g.][]{Cucciati2006, Tasca2009, Scoville2013,  Cucciati2017, Malavasi2017, moutard2018,Laigle2018,PaulinoAfonso2019A&A...630A..57P, Sazonova2020}. 

The morphology-density relation is tightly correlated with colour~\citep[e.g.][]{Poggianti2008,Skibba2009,Bait2017}. 
The existence of such correlations suggests a transformation both in colour and morphology for galaxies in different environments. 
Up to at least $z\sim 2.5$ massive, bright, red, passive early-type galaxies tend to be more clustered and located in denser environments, while the reverse is true for galaxies that have lower mass, are fainter, bluer and star-forming~\citep[e.g.][]{Cooper2007,Dressler1997,Tasca2009,Chuter2011,  Andreon2020A&A...640A..34A, Sazonova2020,Gu2021}. 
However, colour- and morphology-selected samples are not equal,  with spirals being mostly blue galaxies and ellipticals being red, a significant fraction of spiral galaxies are red, and ellipticals blue~\citep[$\sim30\%$, $\sim10\%$, respectively;][and references therein]{Smethurst2022}. 
Moreover, the environmental transformation seems to proceed faster from blue to red than the transformation from spiral to elliptical galaxies~\citep[e.g.][]{Bamford2009, Bolzonella2010}.

The morphology-colour-density relation suggests the key role of the environment in regulating the morphological transformation of galaxies accompanied by quenching of their star formation. 
It was shown in the literature that the number and the total stellar mass, within the passive galaxy population has grown at least by a factor of two from $z\sim1$ to $z\sim0$~\citep[e.g.][]{Bell2004, faber2007}.
During this process, passive galaxies reveal little or no star formation activity~\citep[e.g.][]{siudek17, sanchez2019}. 
This implies that at some point there should occur a transition of galaxies from the blue star-forming to the red passive population via star formation quenching. 
Green (intermediate) galaxies are considered to be in transit between blue star-forming and red passive galaxies, and their limited number has been interpreted as a consequence of a fast process transforming the former to the later ones~\cite[e.g.][Krywult et al. in prep.]{Bell2004,schiminovich07,moutard2016a}, or rejuvenation events~\citep{thomas2010} from recent infall of the gas.
The observed rich panoply of green galaxy morphologies and physical properties suggest the existence of multiple quenching channels~\citep{faber2007, moutard2016a, Pacifici2016b, Siudek2018a,Siudek2018b} demanding both internal and external processes~\citep[e.g.][]{Mahoro2017,Kelvin2018}.

Taking into account all the complex theories listed above, which do not exhaust all possibilities, dividing galaxies into only two (or three) broad galaxy populations may not be sufficient for comprehensive studies of all aspects of environmental influence on galaxy evolution. 
Within one galaxy population, there may exist differences in their properties that can be explained by the various mass assembly pathways of blue star-forming and red passive galaxies. 
For instance, blue galaxies show an intrinsic scatter in the star formation rate-stellar mass relation ~\citep[e.g.][]{Guo2013,Matthee2019}. 
Attributing such a scatter to the existence of different subclasses may reveal the interplay between the various processes that shape how galaxies evolve. 
This thought led to the idea to adopt methods able to select classes of galaxies considering many parameters and not only the morphology and/or the rest-frame colours or their estimated star formation activity. 
In this context, the unsupervised clustering methods can be a valuable approach. 


The VIMOS Public Extragalactic Redshift Survey (VIPERS) provides a unique opportunity for environmental studies at intermediate redshifts ($0.5<z<1.2$), through its combination of large volume coverage (comparable to the 2dFGRS in the local Universe) and a well-defined sample of almost 90\,000 galaxies, whose global properties and local environments are known. Such large samples are needed to permit the division into multiple statistically-representative galaxy sub-populations, that can then be studied and compared. 
In this paper, we use VIPERS to provide a representative sample of galaxies at $z\sim0.7$. 
This unique sample ensures a large number and variety of sources and provides the multi-wavelength coverage needed to quantify the correlation with the environment. 
The multi-wavelength coverage of VIPERS and a wealth of auxiliary data was already used to define 11 different subclasses of galaxies, based on their $UV$-to-$NIR$ colours~\citep{Siudek2018a}, 
 hereafter \citetalias{Siudek2018a} (see Sect.~\ref{sec:types}). 
For each VIPERS galaxy, a well-defined environment estimate is available~\citep{Cucciati2017}, which was already used to study the environmental effects on the galaxy stellar mass function~\citep{davidzon16} and the environment-size relation~\citep{Gargiulo2019}. 

This paper aims to explore the relationships between different galaxy sub-populations and their environment at $z\sim0.7$. Our first aim is to show the fraction-density relation for the full galaxy population seen by VIPERS. 
As the next step, we select (sub)populations promising for more in-depth analysis, which allows breaking the mass-density degeneracy. This paper concentrates on relations of the galaxy types defined based on their rest-frame colours, a definition obtained by an automated classification algorithm that combines 12 rest-frame magnitudes from $UV$ to $NIR$, and local environmental density. 
Consequently, as our classification was based on colours, VIPERS galaxy population-density relation reflects rather the colour than morphology dependence, although the morphological labels are recreated with the accuracy of $0.95$ (see Sec.~\ref{sec:morphology}). 

In this way, we intend to provide both a possibly comprehensive picture of galaxy type-density dependence at $z>0.5$, as a starting point to identify galaxy types whose evolutionary paths can be studied more in detail with this or future data.

The paper is organized in the following way: 
in Sect. \ref{sec:data} we present the VIPERS data sample. 
Section~\ref{sec:3_class_separation} presents the environmental dependence of the fraction of three main galaxy classes (red, green, blue). 
In the following Sec.~\ref{sec:FEMclassenv}, we present results for different subclasses and discuss the environmental dependence of red, green, and blue subclasses in Secs.~\ref{sec:redgalaxies}, \ref{sec:green}, and \ref{sec:starfroming}, respectively. 
The summary is presented in Sect.~\ref{sec:summary}. 
Throughout this paper the cosmological framework with $\Omega_{m}$ = 0.30, $\Omega_{\Lambda}$ = 0.70, and $H_{0}=70$ $\rm{km s^{-1} Mpc^{-1}}$ is assumed.

\section{Data and sample selection}\label{sec:data}

The data used in this paper are taken from the VIMOS Public Extragalactic Redshift Survey~\citep[VIPERS,][]{scodeggio16}. 
VIPERS is a European South Observatory (ESO) Large Program performed by the VIMOS spectrograph~\citep[VIMOS,][]{lefevre03}. 
VIPERS provided spectroscopic redshifts ($z_{spec}$), spectra and full photometrically-selected parent catalogue for 86\,775 galaxies limited to $i_{AB}\leq22.5$ mag over a total area of $\sim$23.5~$\deg^2$ within the W1 (15.7~$\deg^2$) and W4 (7.8~$\deg^2$) fields of the Canada-France-Hawaii Telescope Legacy Survey Wide (CFHTLS-Wide). 
A~detailed description of the survey can be found in~\cite{guzzo}, and \cite{scodeggio16}.
The data reduction pipeline and redshift quality system are described by \cite{garilli14}.	
In addition to the spectroscopic information, the VIPERS data value is enhanced by important ancillary information. In particular, morphological parameters were derived by~\cite{krywult} and local galaxy densities were measured by~\cite{Cucciati2017}.

\subsection{The physical properties}
The physical properties and absolute magnitudes of galaxies were derived from spectral energy distribution (SED) fitting based on $u$, $g$, $r$, $i$, $z$ fluxes from the CFHTLS T0007 release, $FUV$ and $NUV$ measurements from GALEX, near-infrared  $K_{s}$ band from WIRCAM and $K$ from the VIDEO VISTA survey. 
The fitting process was performed with the usage of  $z_{spec}$, with a grid of stellar population models. 
The detailed description of the physical parameters (absolute magnitudes, stellar masses, $\rm{M_{star}}$, and star formation rates, SFR) for the VIPERS sample used in the following analysis can be found in~\cite{moutard2016a}. 
Thanks to the multi-wavelength VIPERS coverage, in the SED fitting we take into account stellar emission, attenuation and re-emission of stellar emission by interstellar dust, which allows us robustly recreate the shapes of SEDs. 
The dependency of the absolute magnitudes on the template library is minimised by using the observed magnitude in the band closest to the redshifted absolute magnitude filter, unless the closest apparent magnitude had an error > 0.3 mag~\citep[see App A.1 in][]{ilbert}. 
Moreover, we are using spectroscopic redshifts, so the physical degeneracies in the redshift-colour space are absent. 

\subsection{The effective radius}
The structural parameters were derived by fitting point spread function (PSF)-convolved S\'ersic profiles to the observed $i$-band CFHTLS-Wide images~\citep{krywult}.  
The fits were performed with GALFIT~\citep{peng2002}, which provides the circularised effective radius, $R_{e}$. 
The accuracy of derived parameters was tested on simulated galaxies based on CFHTLS images, returning the uncertainties in $R_{e}$ measurements at the level of 4.4$\%$ (12$\%$) for 68$\%$ (95$\%$) of VIPERS the sample. 
A~detailed description of the VIPERS morphological parameters can be found in~\cite{krywult}.

\subsection{The environment}\label{sec:env}
As a measure of the environment, we use the VIPERS local density contrast, $\delta$, computed by \cite{Cucciati2017} and defined  as:
\begin{equation}
\delta(RA,DEC,z)=[\rho(RA,DEC,z)-\langle\rho(z)\rangle]/\langle\rho(z)\rangle.
\end{equation}
Here, $\rho(RA,DEC,z)$ is the local density of the tracer centred at the  galaxy $(RA,DEC,z)$, and $\langle\rho(z)\rangle$ is the mean density at the tracer's redshift $z$. 
The density field is computed using a cylinder with a half-length of  $\pm$1\,000 km/sec and the radius equal to the distance of the fifth nearest neighbour defined using a subsample of galaxies that trace the density field (tracers).

Tracers were selected from a volume-limited sample that includes galaxies with  spectroscopic and photometric redshifts. 
Tracers were selected to satisfy $M_{B} \le (-20.4 - z)$, which yields a comoving number density that does not evolve, therefore is not affected by discreteness effects that change with redshift~\citep{Cucciati2017}. 
With this luminosity limit, the tracer (and our) sample is complete up to $z =0.9$. 
A~detailed description of the measurements method and  environment properties can be found in~\cite{Cucciati2014, Cucciati2017}. 

\subsection{Galaxy populations}\label{sec:types}

In~\citetalias{Siudek2018a}, we selected  52\,114 VIPERS objects with the highest confidence ($>90\%$) of redshift measurements and performed their classification into 12 clusters  based on the unsupervised machine learning algorithm known as Fisher Expectation-Maximization~\citep[FEM;][]{Bou2011}. 
The FEM method implements a clustering approach called the Discriminative Latent Mixture (DLM) model. 
It is based on dimensionality reduction as it assumes that data are located in a common low-dimensional latent space. 
The FEM algorithm iteratively estimates both the discriminative subspace and the parameters of the DLM model. 
This ensures that improvements to the estimated parameters of the model are adaptive and that the clustering uses only the most important information encoded within the input features. 
The feature space was defined by spectroscopic redshifts and 12 rest-frame ultraviolet-through-near-infrared magnitudes normalised to the $i$-band magnitude. 
The optimal number of clusters was determined using the Integrated Completed Likelihood criterion~\citep[ICL;][]{BIERNACKI2000}, complemented by the physical interpretation of data flow and the properties of clusters.  
The model revealed substructures in the bimodal colour distribution of galaxies, distinguishing subpopulations of red, green and blue galaxies. 
In~\citetalias{Siudek2018a} we verified the ability of FEM to recognise a naturally defined separation in multi-dimensional space to mirror different mass assembly paths. 
This new categorisation allowed us to distinguish subclasses for red, green and blue galaxy populations and an additional class of broad-line active galactic nuclei (AGNs), which is discussed in Siudek et al in prep. 
In the following part, the definition of the red, green and blue galaxy populations relies on our FEM classification: 
\begin{itemize}
    \item red (subclasses 1--3, C1--3, hereafter), 
    \item green (subclasses 4--6, C4--6, hereafter),
    \item blue (subclasses 7--11, C7--11, hereafter).
\end{itemize}

In particular, subclasses C1--3 host the reddest spheroidal-shape galaxies showing no sign of star formation activity and dominated by old stellar populations (as testified by their strong 4000~\AA~breaks). 
The inhomogeneity among red subclasses is expressed mainly in their stellar content traced by UV colours, revealing star formation activity from young hot stars among C3~\citep[see Fig. 2d in ][and Sec.~\ref{sec:difference_red}]{Siudek2018a}.  
Subclasses C4--6 host intermediate galaxies whose physical properties, such as colours, sSFR, stellar masses, sizes and shapes, are
intermediate relative to red and blue galaxies.
These intermediate galaxies have more concentrated light profiles and lower gas contents than blue galaxies, as observed for green galaxies in the local Universe~\citep{schiminovich07,schawinski}. 
In particular, the colours of galaxies within C4 suggest that they are dust-free transition galaxies, whereas galaxies within C5 correspond to
more dusty galaxies and C6 gathers very dusty star-forming galaxies, potentially edge-on spirals (see Sec.~\ref{sec:green_difference} and~\ref{sec:transition_mass} for more details). 
Subclasses 7–11 contain blue star-forming galaxies. 
The blue cloud of disk-shaped galaxies is actively forming new stars and are populated by young stellar populations (as indicated by the weak 4000~\AA~break) catching them on different stage of their evolution. 
For example, galaxies within C8 are characterised by enhanced star formation activity as indicated by their location on the main sequence plane~\citep[see Fig. 7 in][]{Siudek2018a}. 
Class 11 may consist of low-metallicity galaxies, or AGNs according to its localisation on the AGN diagnostic diagram~\citep[see Fig. 10 in][]{Siudek2018a}. 
The differences between blue subclasses are further discussed in Sec.~\ref{sec:blue_difference}. 

\begin{figure}
	\centerline{\includegraphics[width=0.49\textwidth]{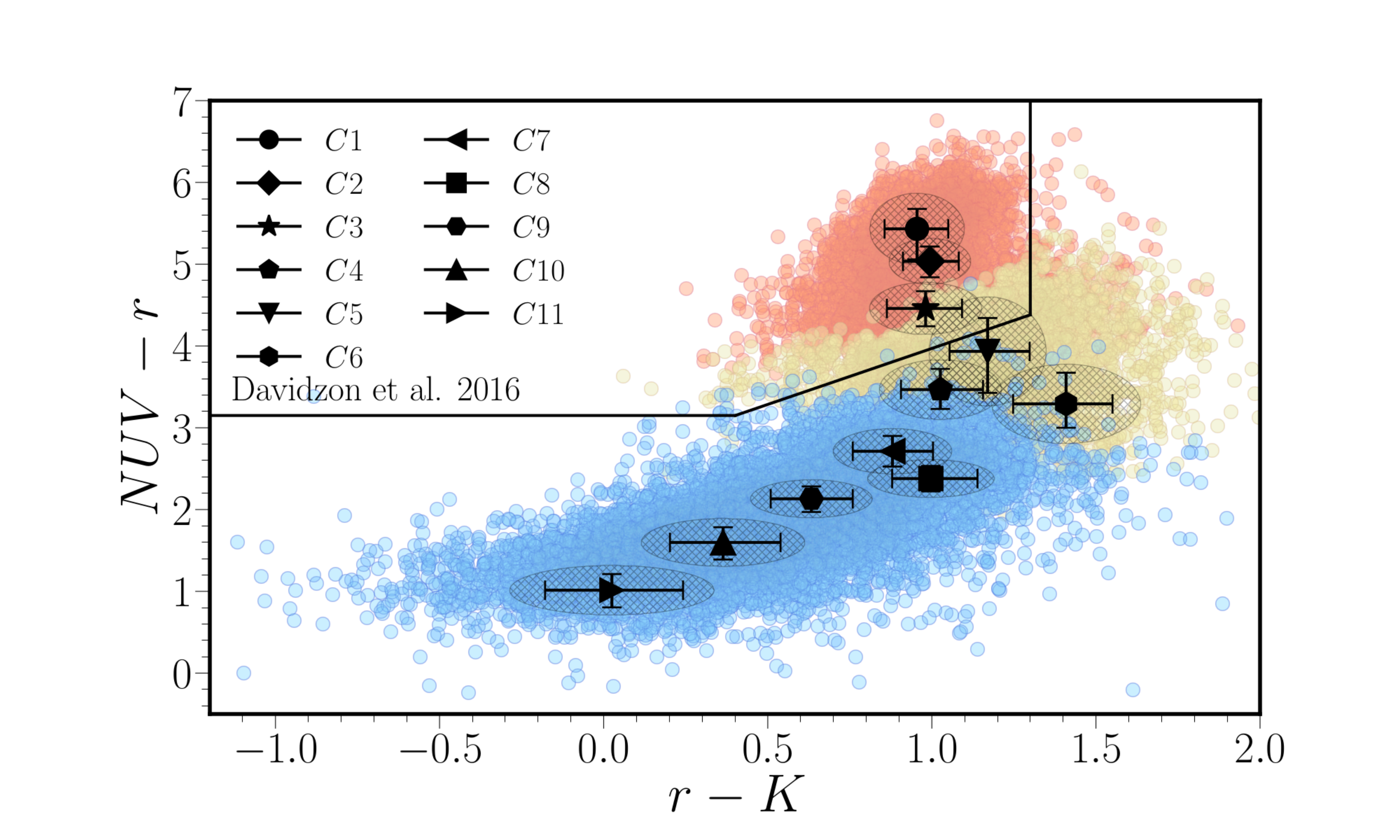}}
	\caption{
	$NUVrK$ diagram of the VIPERS galaxies classified into 11 subclasses. The median colours of subclasses are marked with points. The error bars correspond to the 25th, and 75th percentile range of the galaxy colour distribution. The size of ellipses correspond to the normalised median absolute deviation. Galaxies classified as red (C1--3) are separated from the blue ones (C7--11) by the solid black line as proposed by~\cite{davidzon16}. Green galaxies (C4--6) are located in between red and blue populations. }
	\label{fig:nuvrk}
\end{figure}

Even if such a detailed sub-division is possible only in the multi-dimensional parameter space, it is also well reflected by the $NUV$--$r$ -- $r$--$K$ diagram~\citep[hereafter: $NUVrK$,][]{arnouts}. 
Figure~\ref{fig:nuvrk} shows that the three main classes: red, green, and blue recreate the more standard colour-colour divisions, and they allow for further discrimination of different subclasses.

For instance, the FEM algorithm distinguishes three subclasses of VIPERS green galaxies (C4--6) located in-between the red and blue populations. 
Such a multi-wavelength coverage is necessary to unambiguously identify green valley galaxies~(\citetalias{Siudek2018a}, \citealp{moutard2020}), as different colours yield different bimodalities~\citep[e.g. the blue peak of the $g-r$ distribution occupies the green valley of the $NUVrK$ diagram;][Krywult et al. in prep.]{salim}. 
Only by considering several colours, substructures contributing to the established bimodalities can be uncovered. 

Although many applications of unsupervised techniques to various astrophysical analyses have been reported in recent years~\citep[see][for reviews]{Baron2019, Ball10}, these techniques still pose a number of questions. 
In particular, applying clustering methods raises the question if the cuts in the seemingly continuous distribution of classification feature space reveal genuine clusters, which may be interpreted as distinct galaxy subpopulations. 
In~\citetalias{Siudek2018a} we addressed these doubts by investigating the non-continuous distribution of features that were not used in the classification scheme throughout different clusters (galaxy subclasses). 
In particular, we discussed the comparison between morphological properties: S\'ersic index~\citep{sersic63}; physical properties: $\rm{M_{star}}$, SFR; and spectral properties: 4000\AA~break, $D4000$, equivalent width of OII line, $EW(OII)\lambda3727$, of different subclasses. 
This paper shows how the environment varies among the subclasses of red, green and blue galaxies and how they can be connected to different quenching channels. 

\subsection{The morphological classification}\label{sec:morphology}

Many studies use colours as the proxy for morphological classifications assuming that blue colours indicate late-type (disk) galaxies, while red colours are selecting early-type (elliptical) galaxies~\citep{Strateva2001, Park2005}. 
As the colour transformation might occur on a shorter timescale than the morphological transformation~\citep[e.g.][]{Bamford2009}, the colour-driven classification might be not equal to the morphological one.
In \citetalias{Siudek2018a} we compared FEM classification with morphological one~\citep{krywult} showing that red galaxies are elliptical galaxies, while blue galaxies correspond to disk galaxies. 

To further strengthen the compatibility of our colour-selected galaxy subclasses with morphology, in this Section we compare our classification  with morphological classification from the Dark Energy Survey (DES), which used a supervised deep-learning algorithm~\citep[convolutional neural network;][]{Vega2021}. 
The catalogue was trained on early-type and late-type galaxies with previously known classification presented in~\cite{Dominguez2018} reaching only up to 17.7 mag and their ‘emulated' versions at higher redshifts, to extend the magnitude range of the training sample. 
The test sample (the 'emulated' images with known labels) demonstrated excellent accuracy ($\sim97\%$) up to 21.5 $r$-band  magnitude, suggesting that the network is able to predict correct morphological types based on the features invisible for the human eye. 
The comparison sample of 7\,384 VIPERS galaxies observed up to 21.5 $r$-band magnitude with DES  includes $81\%$ galaxies with secure morphological labels, from which $22\%$ are classified as early-type and $78\%$ as late-type. 
Early-type galaxies are located mainly within C1--4, not contaminating blue subclasses (C7--11). 
At the same time, late-type galaxies dominate in subclasses C7--11, showing only a very small fraction within subclasses C1--3. 
As shown in~\cite{Vega2021}, 89\% of red galaxies (C1--3) are classified as early-type, and 97\% of green/blue galaxies (C4--11) as late-type galaxies, which results in an accuracy classification
score of $\sim0.95$ defined as the fraction of correctly classified
galaxies~\citep[see eq. 5 in][]{Vega2021}.
The intermediate subclasses C4--6 are labelled both as early- and late-type galaxies (with a preference for late-type galaxies). 
This comparison suggests a strong correlation of colours with morphology for our sample~\citep[see Fig.~15 and Sec.~5.3 in ][for more details]{Vega2021}.  

\subsection{The final sample}\label{sec:final_sample}

In the following analysis, we are using the catalogue of 52\,114 representative galaxies presented in \citetalias{Siudek2018a}. After excluding 600 broad-line AGNs (mainly assigned to subclass 12 in~\citetalias{Siudek2018a}), and 3\,966 galaxies with a low probability of belonging to any assigned class, i.e. on the borders between different subclasses see details in~\citetalias{Siudek2018a}) our sample consists of 47\,548 galaxies.
We consider only galaxies with reliable $\delta$ measurements~\citep[i.e. for which at least $60\%$ of the cylinder volume overlaps with the VIPERS survey footprint (gaps and boundaries, see details in][]{Cucciati2017, davidzon16}. 
This criterion limits our final sample to 31\,631 galaxies within the redshift range $0.5<z<0.9$. 
The main physical properties of these galaxies are given in Table~\ref{table:eff}. 
For each galaxy, a statistical weight $w$ accounting for survey incompleteness is provided. 
The selection weight, $w=1/(\rm{TSR}\times\rm{SSR}\times\rm{CSR})$, takes into account three selection functions: the target sampling rate, TSR, the spectroscopic success rate, SSR, and the colour sampling rate, CSR. 
Further details about these criteria are provided in~\citealp{garilli14} and~\citealp{scodeggio16}. 
The selection weight reflects the representatives of a given galaxy with respect to the underlying parent photometric catalogue. 
To account for mass incompleteness introduced by Malmquist bias each VIPERS galaxy is weighted also by the fraction of the volume in which the galaxy would be still observable (using minimal and maximal redshifts at which it could be observed), following the  $1/V_{max}$ method~\citep{Schmidt1968}.

    \renewcommand{\arraystretch}{1.4}

	\begin{table*}
		\centering                         
		\begin{tabular}{p{1.0cm} p{1.0cm} p{1.0cm} p{1.6cm} p{1.6cm} p{1.0cm} p{1.2cm} p{1.2cm} p{1.8cm}}    
			\hline 
			$Cls$ & $N$ & $\delta$ & $\rm{log(sSFR)}$ &  $\rm{log(M_{star})}$ & $R_{e}$  & $n$ & $D4000$ & $FUV-NUV$\\
			\hline 
			\hline
			C1 & 3061 & $1.76_{-1.18}^{+2.35}$ & $-16.88_{-2.16}^{3.55}$ & $10.77_{-0.24}^{0.21}$ &  $2.38_{-0.77}^{+1.33}$ &  $3.37_{-0.97}^{+1.19}$ & $1.76_{-0.10}^{+0.10}$& $3.16_{-0.45}^{0.40}$\\
			C2 & 1637 & $1.77_{-1.11}^{+2.32}$  & $-11.85_{-0.14}^{0.29}$ & $10.81_{-0.21}^{0.21}$ & $2.91_{-1.08}^{+1.36}$ &  $3.29_{-1.05}^{+1.31}$ & $1.74_{-0.11}^{+0.10}$ & $1.60_{-0.22}^{0.24}$\\
			C3 & 2478 & $1.72_{-1.17}^{+2.21}$ & $-11.28_{-0.16}^{0.37}$& $10.81_{-0.23}^{0.23}$ & $3.00_{-1.03}^{+1.29}$ & $3.02_{-0.97}^{+1.49}$ & $1.67_{-0.13}^{+0.12}$ & $0.83_{-0.27}^{0.45}$\\
			\cline{1-9}
			C1--3 & 7176 & $1.75_{-1.16}^{+2.29}$ & $-11.99_{-4.45}^{0.66}$ & $10.79_{-0.22}^{0.22}$ & $2.68_{-0.92}^{+1.39}$ & $3.25_{-1.00}^{+1.32}$ & $1.73_{-0.12}^{+0.10}$ & $1.74_{-0.70}^{1.31}$\\
			\cline{1-9}			
			C4 & 2801 & $1.43_{-1.02}^{+2.00}$ & $-9.73_{-0.63}^{0.44}$ & $10.65_{-0.22}^{0.22}$ & $3.37_{-0.96}^{+1.16}$ & $1.76_{-0.67}^{+1.08}$ & $1.41_{-0.11}^{+0.14}$ & $1.02_{-0.58}^{0.42}$\\
			C5 & 2270 & $1.46_{-1.09}^{+2.09}$ & $-9.57_{-0.17}^{0.33}$ & $10.49_{-0.26}^{0.22}$ & $2.90_{-0.86}^{+1.12}$ & $2.03_{-0.74}^{+1.17}$ & $1.49_{-0.15}^{+0.16}$ & $2.28_{-0.28}^{0.11}$\\
			C6 & 688 & $1.22_{-0.93}^{1.79}$ & $-9.21_{-0.35}^{0.38}$ & $10.50_{-0.20}^{0.21}$ & $3.77_{-1.01}^{+1.26}$ & $1.35_{-0.51}^{+0.85}$ & $1.36_{-0.10}^{+0.11}$ & $0.71_{-0.45}^{0.27}$\\
			\cline{1-9}	
			C4--6 & 5759 & $1.42_{-1.04}^{2.00}$ & $-9.57_{-0.27}^{0.36}$ & $10.57_{-0.24}^{0.23}$ & $3.25_{-0.98}^{+1.27}$ & $1.77_{-0.66}^{+0.98}$ & $1.43_{-0.12}^{+0.16}$& $1.44_{-0.76}^{0.80}$ \\		              
			\cline{1-9}
			C7 & 3281 & $1.22_{-0.96}^{1.67}$ & $-9.23_{-0.49}^{0.44}$ & $10.33_{-0.28}^{0.29}$ & $3.48_{-0.93}^{+1.19}$ & $1.11_{-0.38}^{+0.72}$ &  $1.28_{-0.07}^{+0.08}$ & $0.79_{-0.50}^{+0.55}$\\
			C8 & 1203 & $1.01_{-0.85}^{1.39}$ & $-8.77_{-0.29}^{0.27}$ & $10.08_{-0.21}^{0.20}$ & $3.32_{-0.84}^{+0.93}$ & $0.89_{-0.31}^{+0.66}$ & $1.22_{-0.05}^{+0.06}$ & $0.21_{-0.15}^{+0.16}$\\
			C9 & 3468 & $0.99_{-0.83}^{1.55}$ & $-8.94_{-0.38}^{0.37}$ & $9.88_{-0.23}^{0.25}$ & $3.11_{-0.85}^{+1.09}$ & $0.90_{-0.28}^{+0.62}$ & $1.21_{-0.05}^{+0.06}$ & $0.78_{-0.49}^{+0.28}$\\
			C10 & 9207 & $0.84_{-0.77}^{1.36}$ & $-8.87_{-0.19}^{0.30}$ & $9.56_{-0.19}^{0.20}$ & $2.96_{-0.81}^{+0.90}$ & $0.92_{-0.30}^{+0.60}$ & $1.16_{-0.05}^{+0.06}$ & $0.21_{-0.14}^{+0.26}$\\
			C11 & 1537 & $0.69_{-0.71}^{1.31}$ & $-8.76_{-0.21}^{0.37}$ & $9.22_{-0.15}^{0.17}$ & $2.51_{-0.81}^{+0.96}$ & $1.10_{-0.45}^{+0.80}$ & $1.08_{-0.06}^{+0.07}$ & $0.07_{-0.12}^{+0.19}$\\
			\cline{1-9}	
			C7--11 & 18696 & $0.93_{-0.81}^{1.44}$ & $-8.93_{-0.29}^{0.36}$ & $9.72_{-0.28}^{0.34}$ & $3.06_{-0.84}^{+0.99}$ & $0.94_{-0.30}^{+0.56}$ &  $1.18_{-0.06}^{+0.07}$ & $0.30_{-0.21}^{+0.47}$\\				
		\end{tabular}
		\caption{Physical properties of the final sample. The number of galaxies ($N$), median values of $\delta$, specific star formation rate, $\rm{log(sSFR)}$ $\rm{[yr^{-1}]}$, stellar mass, $\rm{log}(M_{star}/M_{\odot})$,  effective radius, $R_{e}$ [kpc], S\'ersic index, n, $D4000$ and rest-frame $FUV-NUV$ colour for each subclass, $Cls$, are provided. Errors correspond to the differences between median and the 1st, and the 3rd quartile, respectively. 
			 }             
		\label{table:eff}     
	\end{table*}
	
For the parts of the analysis that include the size of the galaxy, a further cut on the $R_{e}$ measurement was introduced, leading to a sample of 25\,555 galaxies with reliable $R_{e}$ estimates (see~\citealp{krywult} for details). 
We have to claim here that the advantage of our studies is the largest galaxy sample over an unprecedented volume with observation of 31\,631 (25\,555 with reliable size measurements) galaxies with spectroscopic redshifts $0.5<z<0.9$.

\section{The main 3-class separation}\label{sec:3_class_separation}

\begin{figure*}
	\centerline{\includegraphics[width=0.99\textwidth]{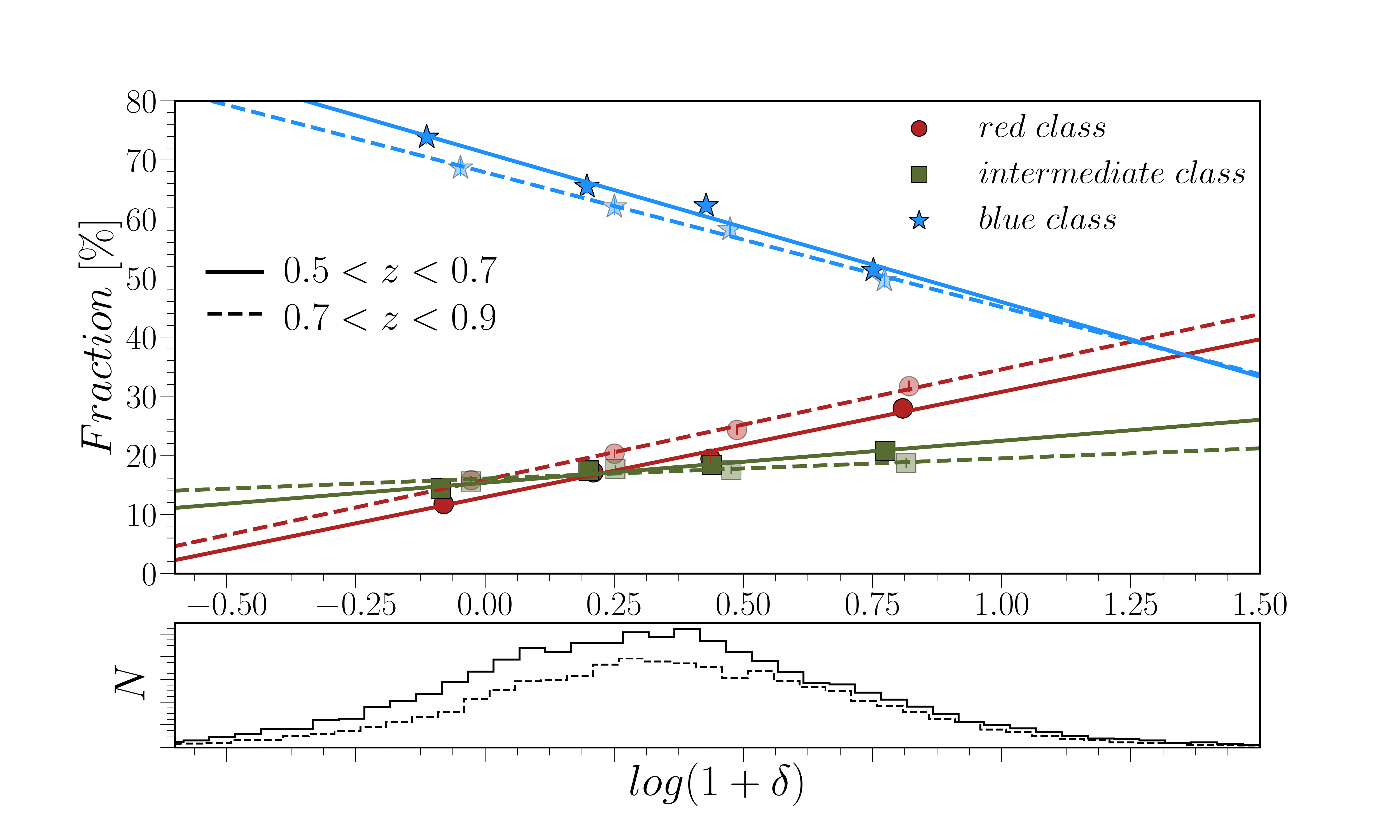}}
	\caption{Upper panel: Fractions of each galaxy VIPERS galaxy population (red, green, and blue class) as a function of $\delta$ weighted using selection, and $1/V_{max}$ weights. Solid lines correspond to the relation found for redshift range $0.5<z<0.7$, whereas dashed lines indicate the dependence of higher-redshift ($0.7<z<0.9$) galaxies on environment. 
	Bottom panel: Distribution of $\delta$ for the final sample of 31\,631 galaxies. Solid and dashed lines correspond to low- and high- redshift bins, respectively. 
	}
	\label{fig:3class_fraction}
\end{figure*}

In the following Sections, we investigate how the fraction of galaxies in different classes vary as a function of the environment.  
For this reason, we identify four $\delta$ bins according to quartiles of the $\rm{log}(1+\delta)$ distribution of the final VIPERS sample. 
The \textit{bottom panel} in Fig.~\ref{fig:3class_fraction} shows the distribution of $\rm{log}(1+\delta)$. 
Since VIPERS is a magnitude-limited survey, we applied corrections (selection and $1/V_{max}$ weights) for survey incompleteness described in Sec.~\ref{sec:final_sample}.
We trace the galaxy population-$\delta$ relation for red (C1--3), green (C4--6) and blue (C7--11) galaxy classes in two redshift bins. 

The \textit{upper panel} in Figure~\ref{fig:3class_fraction} shows how the fractions of red, green and blue galaxies depend on the $\delta$. 
To verify if we see an evolution of the galaxy population-environment relation, we divided the sample into two redshift bins. 
To ensure good statistics we divided the sample at $z=0.7$, resulting in two redshift bins $0.5-0.7$ and $0.7-0.9$, containing 18\,259 and 13\,372 galaxies, respectively. 
As Fig.~\ref{fig:3class_fraction} shows, the trends of red, green and blue galaxies are similar in both our redshift bins, as supported by the agreement of median values in each $\delta$ bin for each class in both redshift bins. 
The choice of the tracer sample (see Sec.~\ref{sec:env}) ensures that densities are not affected by discreteness effects that change with redshift, allowing us to compare the trends in two redshifts bins. 
However, the direct comparison between fractions of different galaxy classes might still be biased to some extent as the blue galaxy sample is predominantly composed of lower-mass galaxies than the red galaxy sample. 
Similarly, the higher/lower redshift samples have different mass distribution, due to VIPERS sample selection, which is addressed by adopting the selection and $1/V_{max}$ weights in our sample.
The possible mass incompleteness problem we partially address in the following Sections by creating a \textit{mass-matched sample}, which ensures the same $\rm{M_{star}}$ distribution for the galaxy subclasses. 

Nevertheless, based on Fig.~\ref{fig:3class_fraction} we can safely conclude that in the whole VIPERS redshift range blue galaxies are more abundant in low-$\delta$ environments, while red galaxies prefer denser regions. 
The blue galaxy fraction steadily decreases from $\sim70\%$ in the lowest $\delta$ bin to $\sim50\%$ at the high-$\delta$ bin. 
This drop is compensated by a corresponding rise in the fraction of green and red galaxies. 
The overall fraction of red galaxies is low but clearly compensates for this trend, increasing their fraction two-fold from $\sim15\%$ to $\sim30\%$ from the low- to high- $\delta$. 
At the same time, the increase of the fraction of green galaxies with $\delta$ is much flatter, going from $\sim15\%$ to $\sim20\%$. 
Our results suggest that up to $z\sim0.9$ red galaxies are preferentially found in denser environments, possibly a consequence of the fact that in denser environments galaxies form earlier and/or stop their star formation earlier.

Using the COSMOS-ACS data in the redshift range $0.0<z<1.2$, \cite{capak2007} found that the fraction of early-type galaxies increases gradually with $\delta$~\citep[especially at $z>0.4$, see Fig.~10 in][]{capak2007}. 
Based on the COSMOS~10k survey \cite{Tasca2009} explored the evolution of the morphology-$\delta$ relation on the sample of 10\,644 galaxies up to redshift $z\sim 1$. 
They found a gradual increasing trend with $\delta$ for early-type E/S0 galaxies at redshift $z<0.5$ and $\rm{log}(M_{star}/M_{\odot})<10.8$ with the flatter slope for more massive galaxies at higher redshift (see Fig.~7 in~\citealp{Tasca2009}). 
\cite{PaulinoAfonso2019A&A...630A..57P} confirmed the existence of a colour-morphology-$\delta$ relation at $z\sim0.84$ based on the sample of $\sim500$ galaxies from the VIMOS Spectroscopic Survey of a Superstructure in COSMOS. 
Their results suggest that environment affects colour as well as structure and morphology.
As the colour-morphology-$\delta$ relation seems to be correlated (see Sect.~\ref{sec:introduction}) and our colour classification follows the morphological types (see Sec.~\ref{sec:morphology}), we claim that our results are in agreement with~\cite{capak2007}, \cite{Tasca2009} and \cite{PaulinoAfonso2019A&A...630A..57P}. 
Our results suggest that the environmental dependence of galaxy morphology is tightly correlated with colour indicating the transformation both in star formation activity and morphology in different environments as it was speculated by~\citealp[e.g.][]{Poggianti2008}, \citealp{Skibba2009}, \citealp{Bait2017}, or \citealp{PaulinoAfonso2019A&A...630A..57P}. 

We show that the general trends with the environment are recovered in the  population of red and blue galaxies beyond the local Universe. 
The fraction of green galaxies shows modest increase ($\sim30\%$) with $\delta$, much weaker than for red galaxies ($\sim100\%$), with a fraction on the level of $15-20\%$. 
The low fraction and flat trend is in agreement with the analysis of the SDSS green galaxies presented in~\cite{Das2021}. 
Those authors found that the fraction  of green galaxies is on the level of $10-20\%$ and it does not depend on the environment. 
Similarly to our sample (see Sec.~\ref{sec:morphology}), the local green galaxy sample consists in majority ($\sim95\%$) of  spiral (late-type) galaxies. 

The modest dependence of the fraction of green galaxies on the $\delta$ means that they are found in both low- and high-$\delta$ environments, suggesting that either external or internal processes may play an important role in regulating star formation in this population. 
In Sec.~\ref{sec:green} we show that their fractions are  dependent on their $\rm{M_{star}}$.

The comparison of trends in two redshift bins allows us to conclude that the galaxy population-$\delta$ relation does not change from redshift $0.5$ up to $0.9$. 
As mentioned before, this comparison might still be affected by sample selection bias, which is partially addressed in the following Sections by creating a \textit{mass-matched sample}. 
The lack of evolution between these two redshift bins indicates that the relation of the galaxy populations (defined by colour) on $\delta$  was already largely in place by $z\sim0.9$. 
This is in agreement  with previous studies showing the existence of such a relation (for galaxies selected by colour or morphology) up to $z\sim0.9$~\citep[e.g.][]{Cooper2007,capak2007, Tasca2009, PaulinoAfonso2019A&A...630A..57P}.
However, a quantitative comparison with other works is not straightforward due to different classifications as well as different environment definitions used in the other surveys. 

\section{The environmental properties of 11 subclasses }\label{sec:FEMclassenv}

While it is largely accepted that different galaxy populations (defined by their colours or  morphologies) preferentially reside in different environments~\citep[e.g.][]{ Dressler1980,balogh2004,Baldry2006,Haines2007,Tasca2009,Cucciati2017}, it has not yet been shown whether this segregation is related to the galaxy subpopulation. 
Different approaches to the galaxy classification were proposed, starting with a visual inspection of small samples to the automatic clustering algorithms dealing with large galaxy surveys~\citep[for review see e.g.][]{FBD15}, but mostly focused on 2(3)-class categorisation.  
Taking advantage of the multi-dimensional classification of VIPERS galaxies (presented in~\citetalias{Siudek2018a}), in the following Sections we show that galaxy sub-populations (subclasses) show different environmental properties to each other.

\begin{figure}
	\centerline{\includegraphics[width=0.49\textwidth]{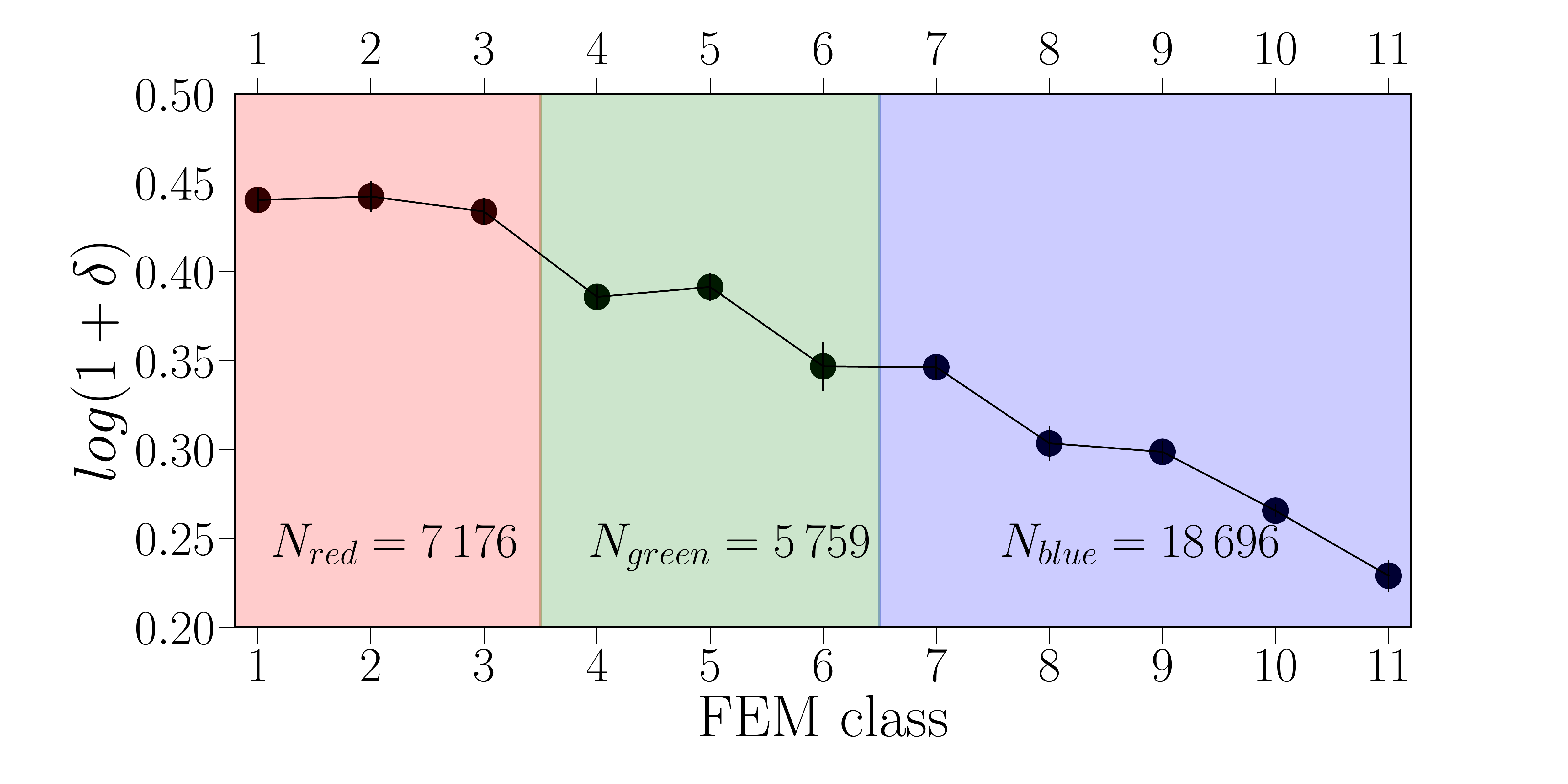}}
	\caption{Median $\delta$ for red (C1--3), green (C4--6) and blue (C7--11) galaxies are shown in red, green, and blue, respectively. The error bars correspond to the median absolute deviation. The number of galaxies classified as red, green, and blue are given in the bottom. }
	\label{fig:11class_density}
\end{figure}

In Fig.~\ref{fig:11class_density} we show the median $\delta$ as a function of the subclass. 
As is clearly visible the subclasses are strongly correlated with the $\delta$. 
The general trend indicates that the median $\delta$ is decreasing towards the bluest galaxies (i.e. with an increasing number of the subclass). 
Subclasses containing the red galaxies reside in the densest environments, while blue galaxies are found in low-$\delta$ regions with green galaxies located in-between. 
Moreover, we can see that the changes in $\rm{log}(1+\delta)$ occur also between different subclasses of green and blue galaxy populations. 
The median values of $\delta$ together with the main physical properties for 11 subclasses are given in Table~\ref{table:eff} and the distributions of $\rm{log}(1+\delta)$ are shown in App.~\ref{app:delta_distribution}.

In the following Sections, we will inspect the fraction of galaxies  for each subclass of red/green/blue population as a function of the environment. 

If not specified differently, the $\delta$ bins correspond to the $\delta$ distribution of the whole VIPERS sample and the fraction is given as the number of galaxies in each $\delta$ bin over all galaxies within each subclass, i.e. for each subclass, the fractions in $\delta$ bins sum up to 100\%. 
This allows us to compare directly the slopes between different subclasses.

\section{Red galaxies}\label{sec:redgalaxies}

\begin{figure}
	\centerline{
	\includegraphics[width=0.49\textwidth]{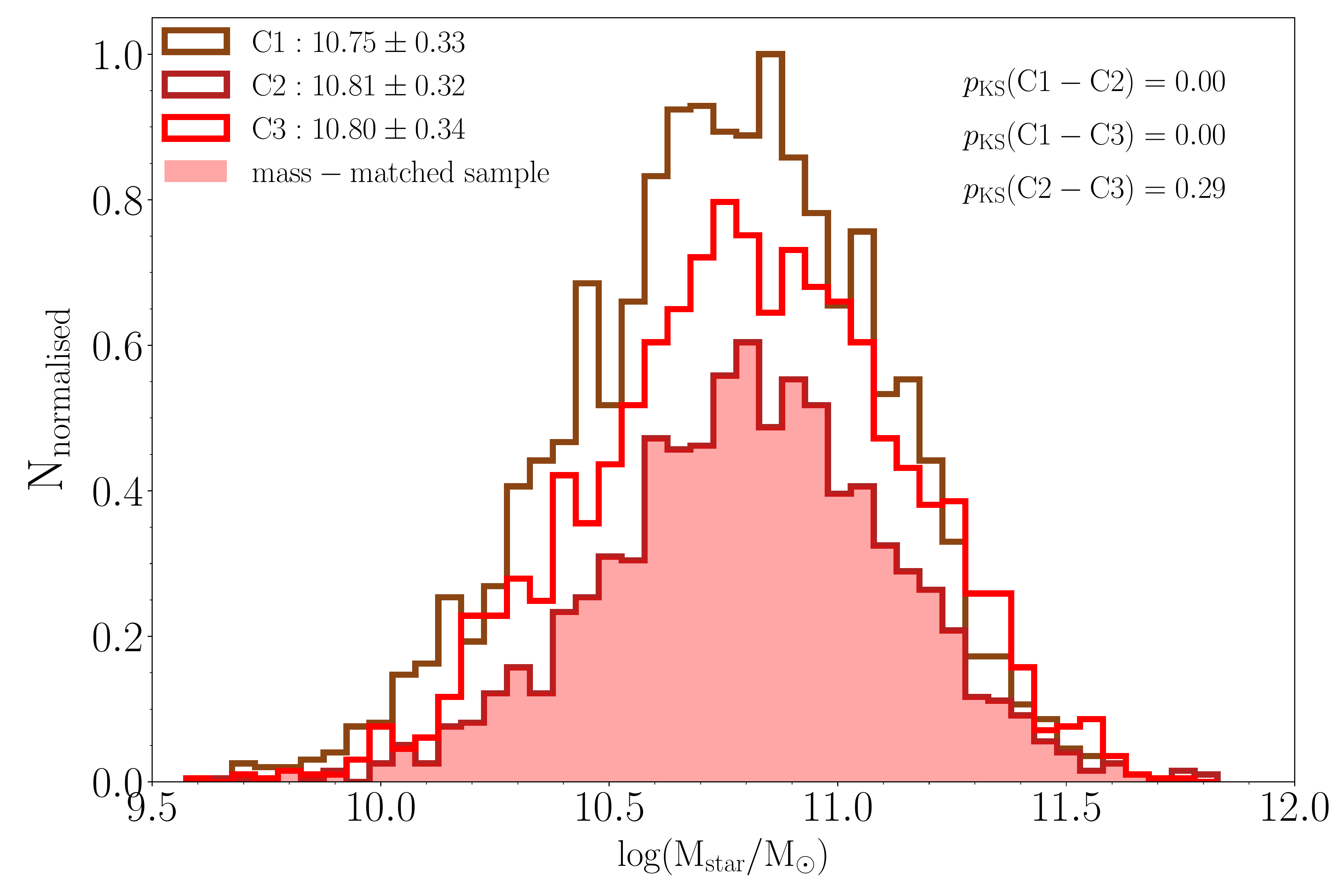}}
	\caption{$\rm{M_{star}}$ distributions of red subclasses C1--3 normalised to the maximum value of the histograms. The mean and standard deviation is shown in the upper left. The KS probability is displayed in the upper right. The red filled histogram indicates the $\rm{M_{star}}$ distribution for the  \textit{mass-matched sample}, which is the same for C1--3.}  
	\label{fig:redhisto}
\end{figure}

We use our VIPERS' sample of 7\,176 red passive galaxies at redshifts between 0.5 and 0.9 to inspect their environmental dependence. 
First, we compare $\rm{M_{star}}$ distribution within C1--3, as the observed environmental trends may be driven by the difference in their masses. 
As it is clear from Fig.~\ref{fig:redhisto} the $\rm{M_{star}}$ distribution is similar for C1--3, as supported also by their median (see Tab.~\ref{table:eff}) and mean (reported in the legend) values ($<0.5\sigma$).
However, there is a hint of C1 inhabiting slightly less massive galaxies (by 0.04-0.06 dex based on their mean/median values). 

To quantify the comparison of their $\rm{M_{star}}$ distribution for three red subclasses, we calculated the two-sample Kolmogorov-Smirnov (KS) null probability for each pair
of C1--3 $\rm{M_{star}}$ distributions, where small values indicate that the two distributions in question are probably not from the same underlying distribution. 
The outcome of the KS test for C1 ($p_{KS}=0.0007, 0.0003$ for sample pairs: (C1, C2) and (C1, C3), respectively) indicates that with high probability C1 and 2 or 3 were not drawn from the same parent population. 
However, we cannot reject the hypothesis that the mass distributions of the C2 and 3 are the same ($p_{KS}=0.29$ for sample pair (C2, C3)). 
Therefore, we create a \textit{mass-matched sample} by drawing from C1 and C3 a sample that reproduces the same histogram in the $\rm{M_{star}}$ as C2 (the sample with the lowest number of galaxies in each mass bin) providing that the total (overall) mass distribution  is the same. 
The \textit{mass-matched sample} was constructed by separating galaxies into narrow $\rm{M_{star}}$ bins of 0.05 dex width and randomly extracting the same number of galaxies as in the smallest bin. 

\subsection{The fraction-$\delta$ relation}\label{sec:red_fraction}

\begin{figure}
	\centerline{\includegraphics[width=0.49\textwidth]{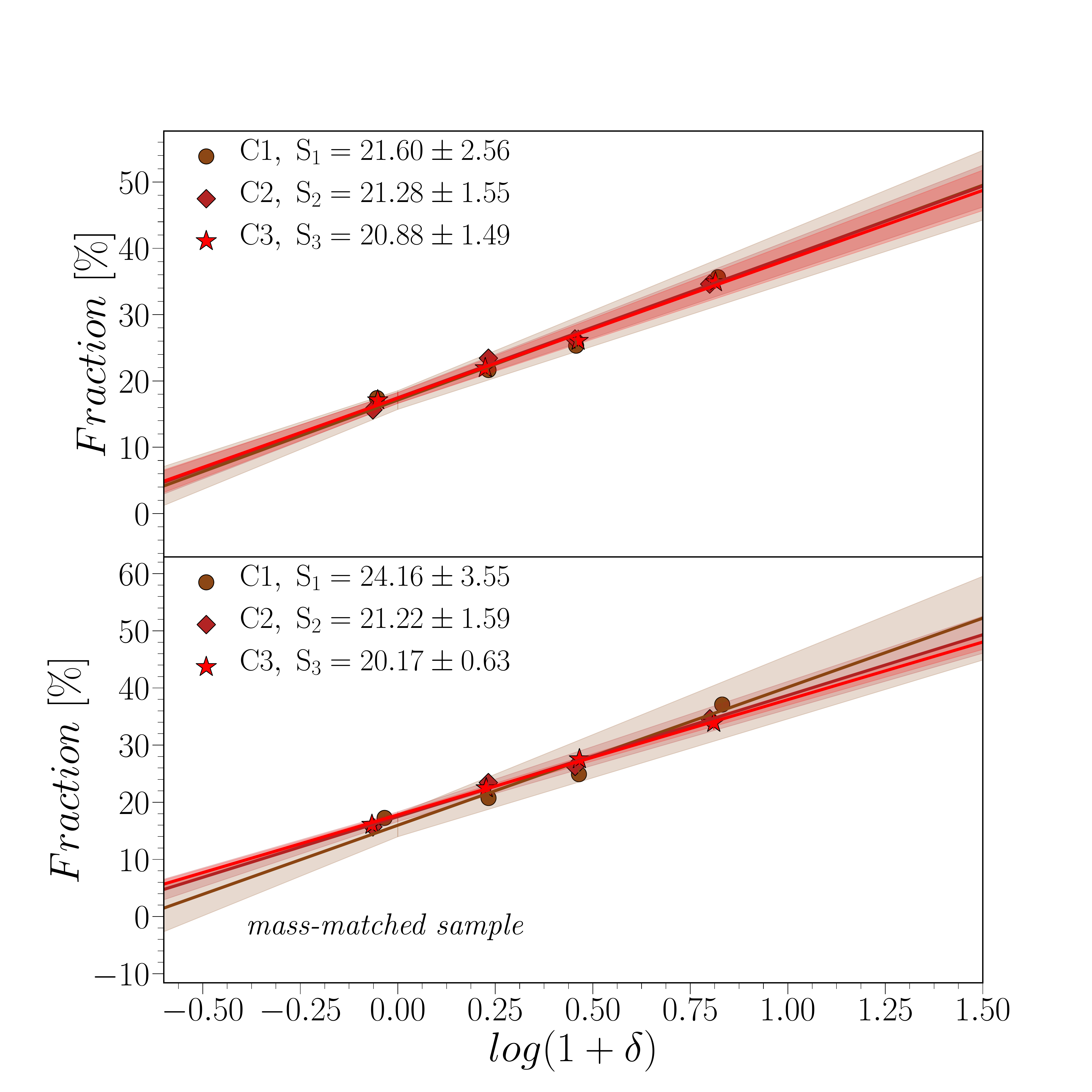}}	
	\caption{Upper panel: Fraction of 3\,061, 1\,637, 2\,478 red galaxies within C1, 2, and 3, respectively, as a function of the $\delta$. The $\delta$-bins corresponds to the quartiles of the $\delta$ distribution for the whole sample, and the fraction are normalised to the total number of galaxies in the same subclass, i.e. for each subclass the fractions in $\delta$ bins sum up to 100\%. The solid line corresponds to the weighted fit. Shaded stripes around lines display $1\sigma$ of the fit. The error bars correspond to the median absolute deviation on this and all the remaining plots. The slope of the fit is given in the legend. Bottom panel: As for the top panel, but for \textit{mass-matched sample}. The $\rm{M_{star}}$ distribution for each red subclass is the same. }
	\label{fig:redfraction}
\end{figure}

The relation between fraction of the galaxy subclass and the environment for C1--3 is shown in Fig.~\ref{fig:redfraction}. 
Galaxies within C1--3 follow the general trend found for red galaxies (see Fig.~\ref{fig:3class_fraction}), with all three red subclasses showing a preference to be in a higher $\delta$ environment. 
This preference is similar in strength for all three subclasses, showing a similar increase by factor of 2 between their number in the first and last $\delta$ bin (the fraction rises from $\sim15\%$ to $\sim35\%$). 
This dependence of their fraction on the environment seem not to be caused by the difference in the $\rm{M_{star}}$ distribution within C1--3 as \textit{mass-matched sample} shows similar trend (see \textit{bottom panel} in Fig.~\ref{fig:redfraction}).
The slopes of the fraction-$\delta$ relation are similar for C1--3 (within $1\sigma$), whether we consider \textit{mass-matched sample} or not. 
Our trends of the red fraction-$\delta$ relation (with an average slope of $21.2\pm1.9$) are similar to the ones, S=$19.4\pm9.8$,  found by~\cite{Cooper2007} based on the analysis of 76 red galaxies at $0.4<z<0.75$ from the DEEP2 galaxy survey. 

\subsection{The size-$\delta$ relation} \label{sec:red_radius}

\begin{figure}
	\centerline{\includegraphics[width=0.49\textwidth]{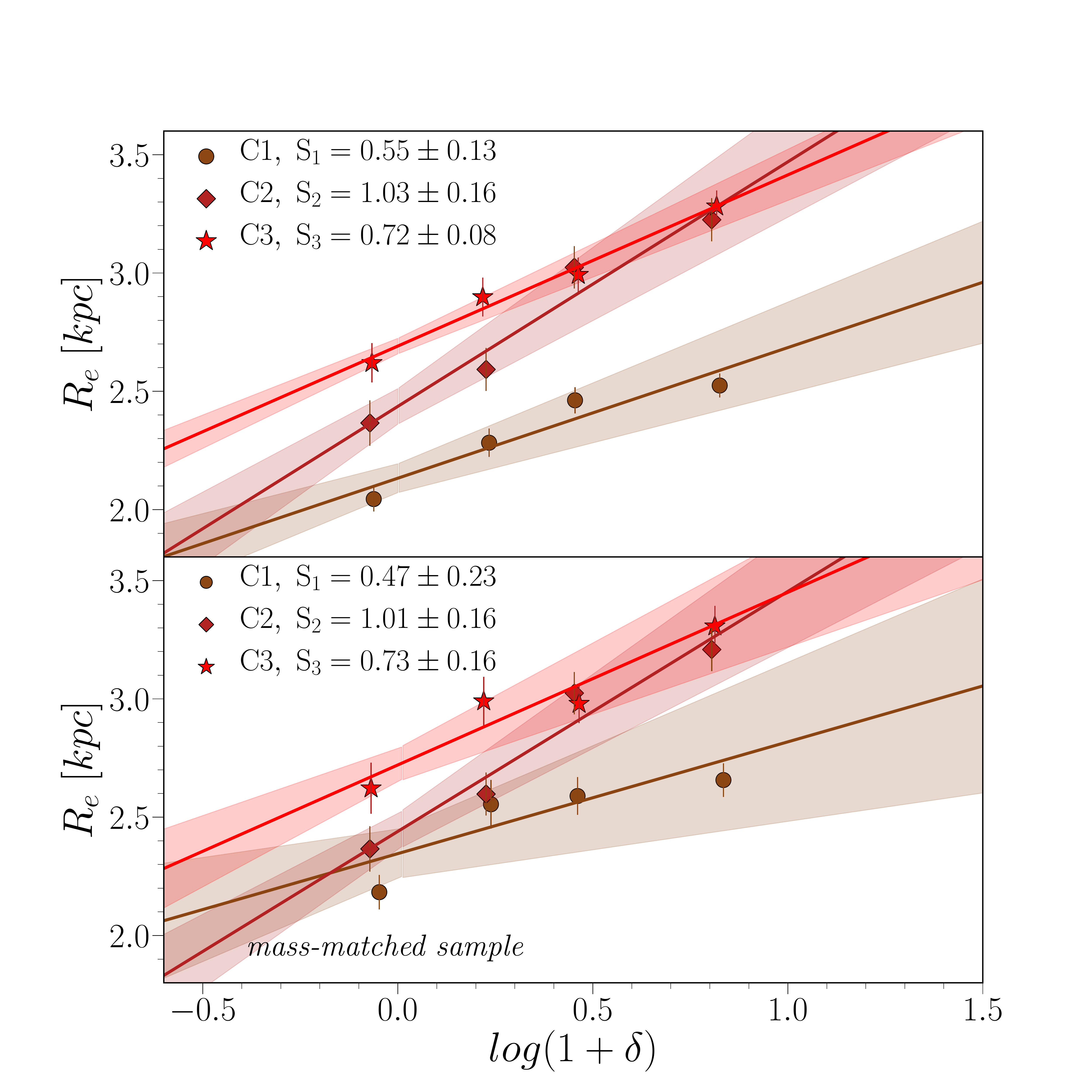}}
	\caption{Upper panel: $R_{e}$-$\delta$ relation for C1, 2, and 3. The solid line corresponds to the weighted fit. Shaded stripes around lines display $1\sigma$ of the fit. The slope of the fit is given in the legend. 	 Bottom panel: As for the top panel, but for the \textit{mass-matched sample}. The $\rm{M_{star}}$ distribution for each red subclass is the same.}
	\label{fig:RED_ReDensity}
\end{figure} 

Before analysing the size-$\delta$ relation, we note that C1, 2, and 3 form a sequence of increasing galaxy size, with C1 gathering the smallest one (see Tab.~\ref{table:eff}). 
They are $\sim$20\% smaller than galaxies within C2 and 3. 
Even though the size was not used as a classification feature, the FEM algorithm used in \citetalias{Siudek2018a} was able to distinguish the cluster (galaxy subclass), which shows differences also in this property. 
This suggests that the  information about size is incorporated in the galaxy SED and confirms the usefulness of the unsupervised multi-wavelength approach to galaxy classifications. 

The dependence of the $R_{e}$ on the $\delta$ for C1--3 is shown in Fig.~\ref{fig:RED_ReDensity}. 
All three subclasses follow the trend of the increasing $R_{e}$ with denser environments. However, the trend is stronger for galaxies within C2 than for the ones in C1 and 3, with C1 showing the weakest dependence of their sizes on the environment. 
Our findings are independent of whether we use the \textit{mass-matched sample} or not. 
This means that even when excluding from C1, the less massive galaxies that are overabundant in low-$\delta$ environments and usually smaller~\citep{Gargiulo2019}, it does not affect the size-$\delta$ relation. 

The upward trend of size-$\delta$ correlation is in agreement with other works~\citep[e.g.][]{Cooper:2012} and supports the scenario in which the evolution of early-type galaxies at $z<2$ is triggered in high-$\delta$ environments through dry minor mergers. 
\cite{capak2007} using data from DEEP2 and DEEP3 surveys found a similar relation between galaxy size and environment at $z\sim0.75$, showing that massive early-type galaxies in denser environments are $25\%$ larger than their equal-mass counterparts in lower-$\delta$ environments. 
This is consistent with our results as on average red galaxies in the high-$\delta$ environment are larger by $28\%$ than the ones in low-$\delta$ environments for the \textit{mass-matched sample}.

\subsection{Differences between red subclasses 1--3}\label{sec:difference_red}

C1 consists of the reddest, most quiescent, the lowest mass and the smallest galaxies from all three red subclasses (see Tab.~\ref{table:eff} and \citetalias{Siudek2018a} for more detailed description).
Their $R_{e}$ have a less pronounced dependence on the environment with respect to galaxies within C2, and 3. 
We claim that dry mergers are not the main driver of the $\rm{M_{star}}$ accretion of galaxies within C1, in agreement with the analysis of VIPERS massive passive galaxies by~\cite{Gargiulo2019}, who found no correlation between the  surface mean stellar mass density and environment for $\sim900$ massive ($\rm{log}(M_{star}/M_{\odot})>11$) passive VIPERS galaxies. 

Furthermore,  C1 may catch a part of the population of red ultra-compact massive galaxies, known as red nuggets.
In Lisiecki et al. submitted, we found 77 red nuggets, of which 49 are found among subclass C1 and the rest is distributed among C2--C5. 
The catalogue of VIPERS ultra-compact massive red galaxies is publicly available (Lisiecki et al. submitted) and their environmental dependence is presented in Siudek et al. in prep. 

In~\citetalias{Siudek2018a} we showed that C2 gathers red galaxies with similar optical colours, the strength of the $D4000$ and S\'ersic indices as for galaxies within C1. 
The main difference in properties of galaxies within C2 in respect to the ones in C1 is in the level of recent star formation activity mirrored in higher specific star formation rate, sSFR, and bluer $FUV-NUV$ colours (see Tab.~\ref{table:eff}). 
Here, we show that galaxies within C2 show the strongest correlation of their $R_e$ with the environment (see Fig.~\ref{fig:RED_ReDensity}). 
Their sizes grow with $\delta$ by $36\pm9\%$ for the \textit{mass-matched sample} clearly supporting the hierarchical scenario. 
It can be explained that galaxies in dense regions evolve faster than galaxies in the field since in higher densities more mergers occur per unit time~\cite[e.g.][]{deLucia:2004ApJ,Andreon:2018}. 
This implies that mergers that are efficient in increasing sizes were relevant in the evolution of galaxies within C2.

For galaxies within C3 the $D4000$ strength, optical colours or S\'ersic indices are on a similar level as for galaxies of the two remaining red subclasses. 
However, galaxies within C3 show the highest sSFR and the bluest $FUV-NUV$ colours 
among red subclasses (see Tab.~\ref{table:eff}), which may indicate that the star formation activity contributes more to C3 than to the C1 and 2 (see~\citetalias{Siudek2018a} for details). 
C3 tends to gather the largest red galaxies, especially in low-$\delta$ environments in respect to C1 and 2 (see Fig.~\ref{fig:RED_ReDensity} and Tab.~\ref{table:eff}). 
The correlation of properties of galaxies within C3 with the environment is in-between the ones found for C1 and 2. 
This may imply that dry mergers occur among those galaxies but it is not the only or the main ingredient of their star formation history, SFH.

\section{Green galaxies}\label{sec:green}

\begin{figure}
	\centerline{
	\includegraphics[width=0.49\textwidth]{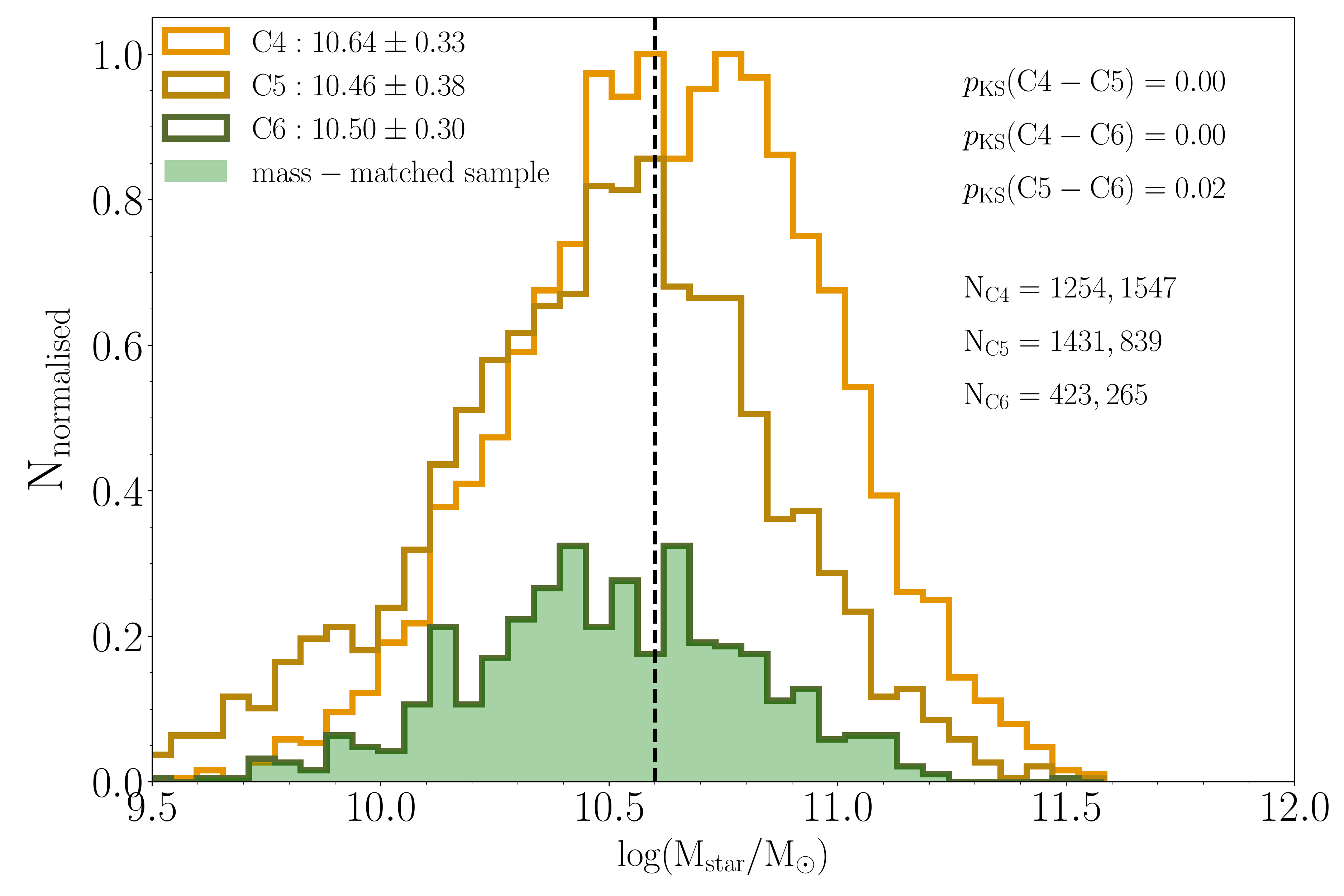}}
	\caption{$\rm{M_{star}}$ distributions of green subclasses  C4--6 normalised to the maximum value of the histograms. The mean and standard deviation is shown in the upper left. The KS probability is displayed in the upper right. The green filled histogram indicates the $\rm{M_{star}}$ distribution for the \textit{mass-matched sample} and corresponds to the area below the distributions for all three subclasses. The vertical line corresponds to the transition mass at $z\sim0.7$. The number of galaxies in low- and high-mass samples, respectively for each subclass are given in the centre-right.} 
	\label{fig:green_massdistribution}
\end{figure}

In this Section, using a sample of 5\,759 green galaxies at redshift $0.5<z<0.9$, we investigate their environmental properties and dependence on the $\rm{M_{star}}$. 
Similarly to red galaxies (see Sec.~\ref{sec:redgalaxies}) we first verified the difference in $\rm{M_{star}}$ distributions within C4--6, which seem to be similar for each green subclass as shown in Fig.~\ref{fig:green_massdistribution} and supported also by their median (see Tab.~\ref{table:eff}) and mean values (see the legend). 
Their $\rm{M_{star}}$ are in agreement within $0.5\sigma$, however, their mean and median values form a sequence of decreasing values which underline different distributions confirmed by KS test (with $p\sim0$) indicating that with high probability C4, 5, and 6 were not drawn from the same parent sample. 
To account for the differences in mass distributions, we create a  \textit{mass-matched sample} following the methodology introduced for red galaxies (see Sec.~\ref{sec:redgalaxies}). 

\subsection{The fraction-$\delta$ relation}\label{sec:green_fraction}

\begin{figure}
	\centerline{\includegraphics[width=0.49\textwidth]{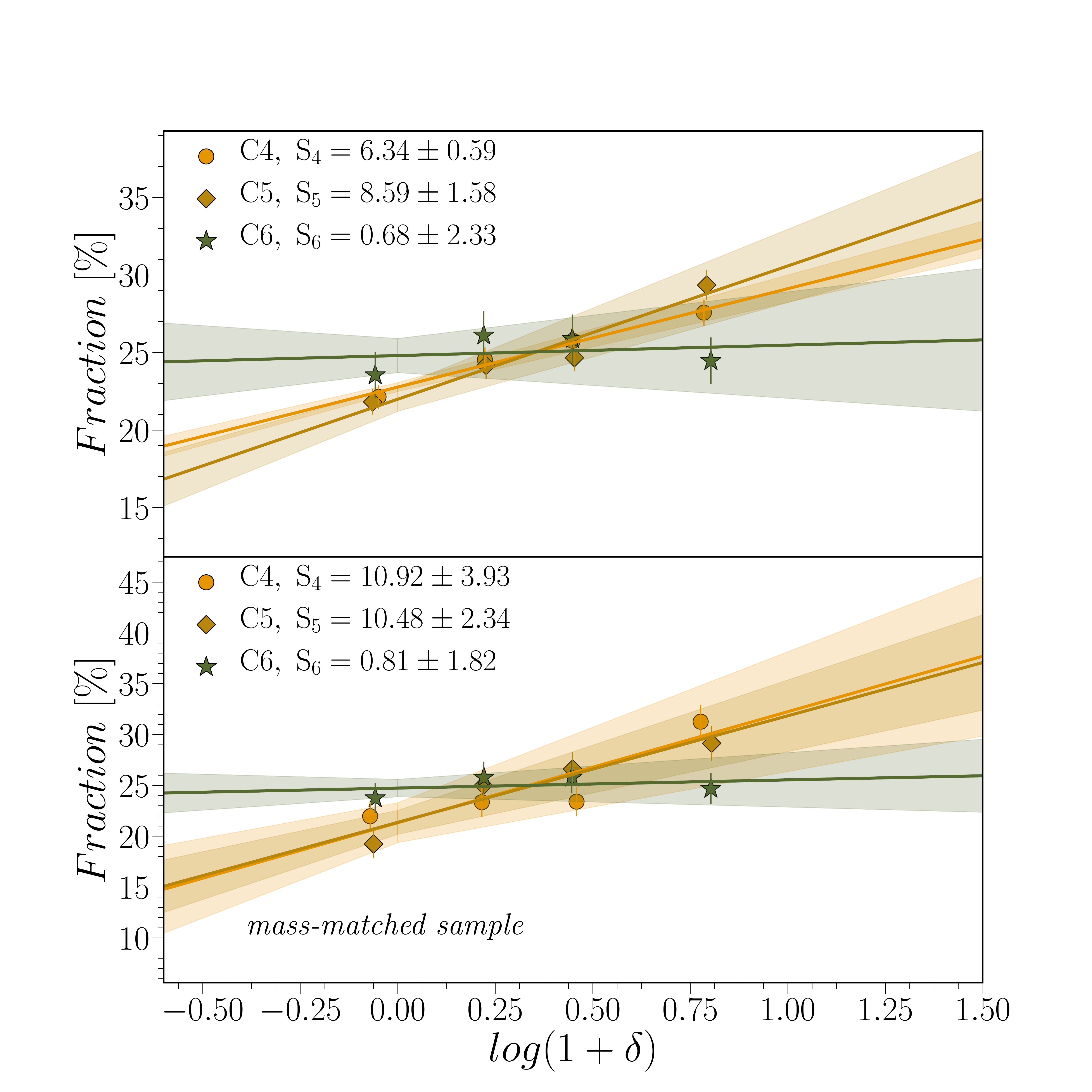}}	
	\caption{Upper panel: Fraction of 2\,801, 2\,270, 688 green galaxies within C4, 5, and 6, respectively, as a function of the $\delta$. The $\delta$-bins corresponds to the quartiles of the $\delta$ distribution for the whole sample, and the fractions are normalised to the total number of galaxies in the same subclass, i.e. for each subclass the fractions in $\delta$ bins sum up to 100\%. The solid line corresponds to the weighted fit. Shaded stripes around lines display $1\sigma$ of the fit. The slope of the fit is given in the legend. Bottom panel: As for the top panel, but for \textit{mass-matched sample}. The $\rm{M_{star}}$ distribution for each green subclass is the same.}
	\label{fig:green_fraction}
\end{figure}

As is clear from Fig.~\ref{fig:3class_fraction}, the fraction of green galaxies is low, showing a small increase towards denser environments. 
In Fig.~\ref{fig:green_fraction} we plot the fraction-$\delta$ relation for the green subclasses C4--6. 
We can see that the trend observed in Fig.~\ref{fig:3class_fraction} is driven by C4 and~5. 
The trend of the relation for C6 is flat ($0.8\pm1.8\%$ for \textit{mass-matched sample}), whereas for C4 and 5 is significantly sharper ($10.7\pm3.1\%$), however still weaker by a factor of two than the one found for red subclasses ($21.9\pm1.9\%$ for \textit{mass-matched sample}).
The fraction-$\delta$ relation seems to be independent of differences in the $\rm{M_{star}}$ distributions within C4--6, as the \textit{mass-matched sample} shows the same trend (see \textit{bottom panel} in Fig.~\ref{fig:green_fraction}).

\subsection{The size-$\delta$ relation}\label{sec:green_fraction}

\begin{figure}
	\centerline{\includegraphics[width=0.49\textwidth]{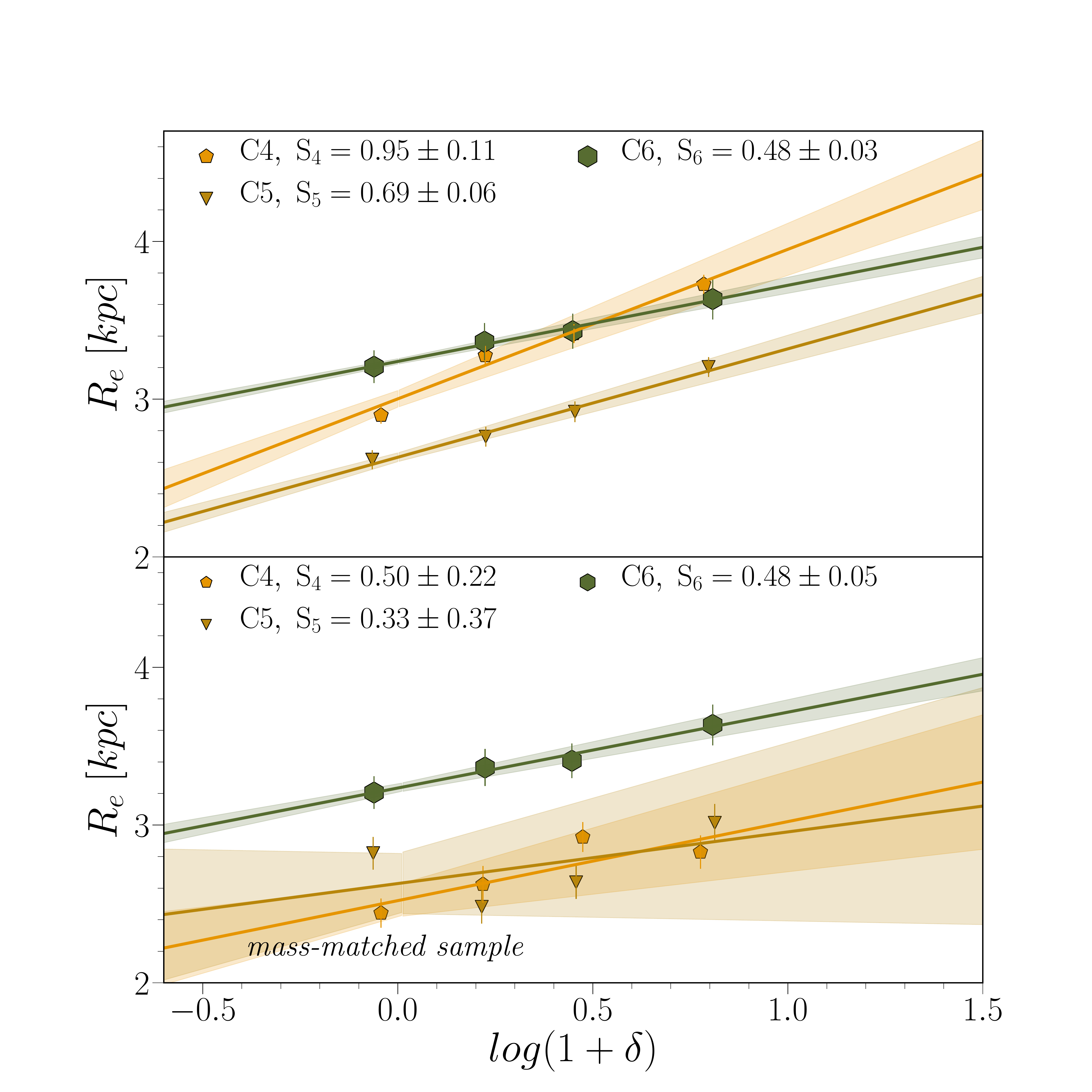}}
	\caption{Upper panel: $R_{e}$-$\delta$ relation for three green galaxy subclasses C4--6. The solid line corresponds to the weighted fit. Shaded stripes around lines display the $1\sigma$ uncertainty of the fit. Bottom panel: As for the top panel, but for the \textit{mass-matched sample}.  The $\rm{M_{star}}$ distribution for each green subclass is the same. }
	\label{fig:green_re}
\end{figure}

Galaxies within C6 are the largest ones, while the ones within C5 are the smallest, placing C4 galaxies in-between (see Tab.~\ref{table:eff}). 
The median size of the galaxies within C5 is similar to the ones within red C2--3, while the ones in C6 are larger by $\sim36\%$ than the ones in the red population and even larger than the ones in the blue population (by $\sim23\%$).
Although the sizes of green galaxies within the three subclasses are different, the dependence of their sizes on the environment is similar. 
The size-$\delta$ relation is shown in Fig.~\ref{fig:green_re}. 
All three green subclasses follow the trend of the mildly increasing $R_{e}$ with $\delta$ ($0.7\pm0.1\%$), which is similar to the trend found for red galaxies ($0.8\pm0.1\%$). 
The size dependence for all three subclasses is in agreement within $1\sigma$. 
However, their size is dependent on their $\rm{M_{star}}$, as the more massive galaxies are larger (for the \textit{mass-matched sample}, which removes the excess of high-mass galaxies, $R_{e}$ is smaller). 
For the \textit{mass-matched sample} for all three subclasses, the size weakly ($\sim7-15\%$) rise from low- to high-densities with a similar strength ($\sim0.5\sigma$).

\subsection{Differences between green subclasses}\label{sec:green_difference}

Galaxies within C6 differ from the ones in C4--5, as they are the largest galaxies within our sample ($R_{e}\sim3.77$~kpc, see Tab.~\ref{table:eff}), including the blue subclasses. 
At the same time, they are relatively massive with reduced star formation (see Tab.~\ref{table:eff}). 
Its location on the $NUVrK$ diagram (see Fig.~\ref{fig:nuvrk}) in  particular is consistent with previous descriptions of the edge-on galaxies~\citep{arnouts, moutard2016b}. 
Their red $r-K$ colours may be a consequence of dust within the disks or their high inclination. 
The presence of the prominent disk also may explain their low S\'ersic indices (with a median of $n=1.35$) in comparison with these of the red population (n$>$3, see Tab.~\ref{table:eff}). 
This suggests that C6 contains spiral galaxies, possibly including dust reddened candidates. 
Those galaxies occupy less dense environments (with respect to red galaxies and remaining green subclasses, see Tab.~\ref{table:eff}),  and their fraction does not depend on the environment suggesting that their evolution is dominated rather by internal processes, such as mass quenching, in which stellar/AGN feedback removes gas from galaxies, while bulges stabilise their disks against collapse inhibiting star formation. 

The dominance of galaxies within C4--5 in denser environments suggests that those galaxies are on a different evolutionary pathway, preferentially dominated by external processes moving these galaxies onto the red sequence over time.

\begin{figure}
	\centerline{\includegraphics[width=0.49\textwidth]{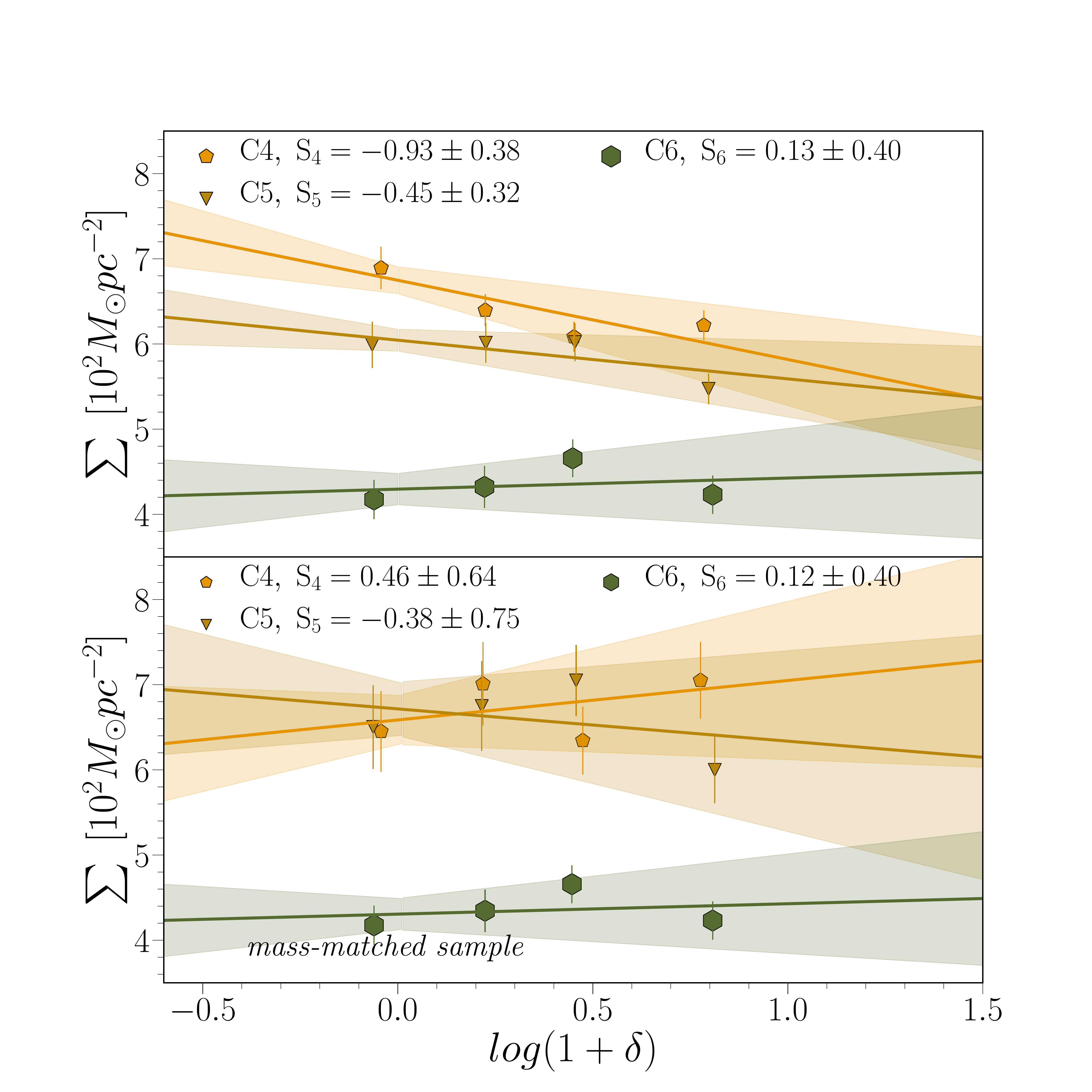}}
	\caption{Upper panel:$\Sigma$-$\delta$ relation for three green galaxy subclasses C4--6. The solid line corresponds to the weighted fit. Shaded stripes around lines display the $1\sigma$ uncertainty of the fit. Bottom panel: As for the top panel, but for the \textit{mass-matched sample}. The $\rm{M_{star}}$ distribution for each green subclass is the same. }
	\label{fig:green_sigma}
\end{figure} 

\subsection{The transition mass}\label{sec:transition_mass}

The green subclasses mostly contain galaxies with $\rm{M_{star}}$ close to the so-called transition mass~\citep[($\log (\rm{M_{cross}/M}_{\odot})=10.6$ 
at $z\sim0.7$;][see Fig.~\ref{fig:green_massdistribution}]{davidzon13}, where the mass functions of the star-forming and passive galaxies cross (above the transition mass the red population dominates, while below that mass, the blue cloud is the most numerous population). 
This sharp transition in $\rm{M_{star}}$ reflects the differences in the evolution of low- and high-mass galaxies~\citep{Haines2007}. 
Low-mass galaxies with shallower potential wells are characterised by more extended star formation histories and longer gas-consumption timescales~\citep{vanzee2001}, while massive galaxies with deeper potential wells consume their gas in a short burst~\citep[$<2$ Gyr;][]{chiosi}. 
In particular, this transition in galaxy properties with mass may be explained by the transition from cold to hot gas accretion modes when galaxy haloes reach a mass of $\sim10^{12}$~\citep{Dekel2006}.

 \begin{figure*}
	\centerline{
	\includegraphics[width=0.99\textwidth]{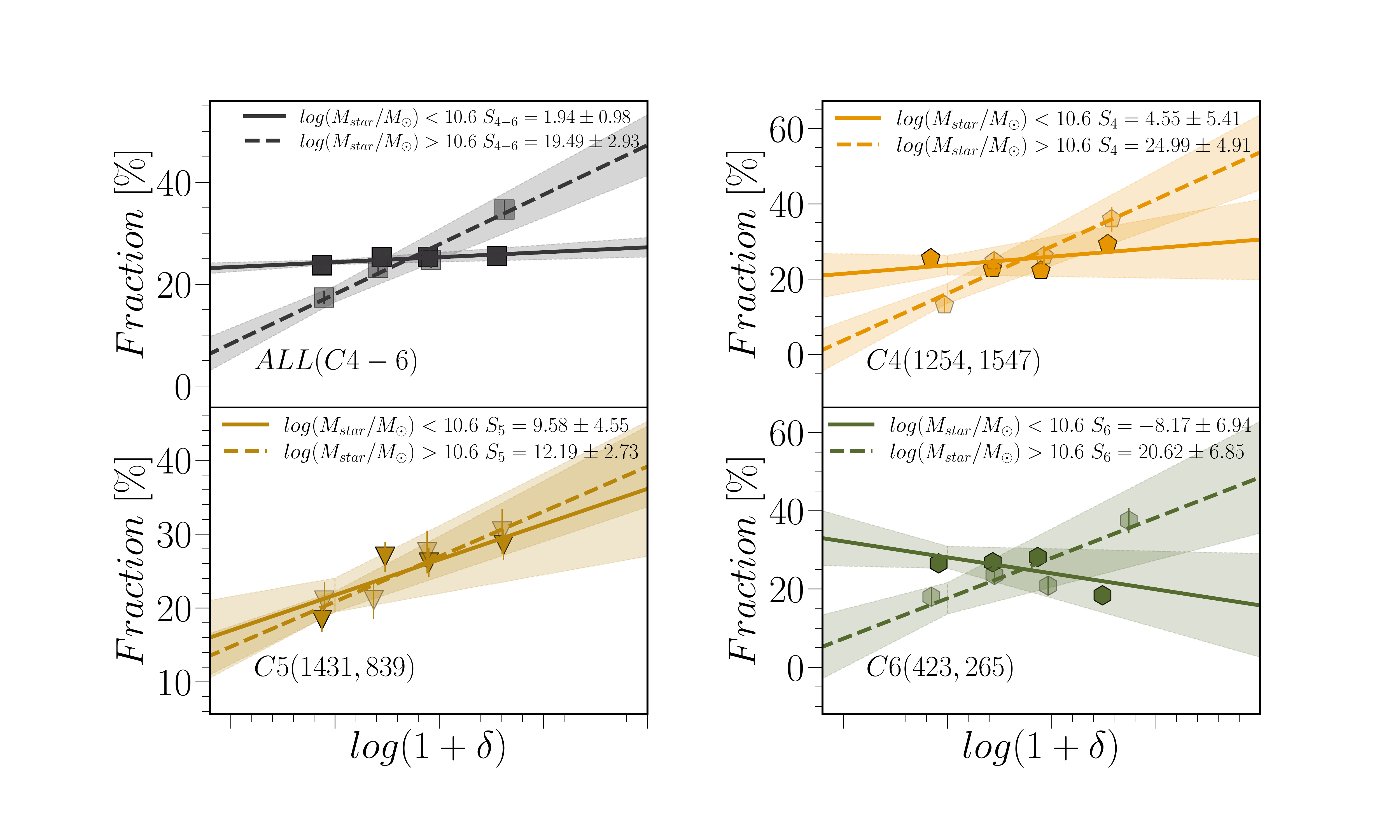}}
	\caption{In each panel, the fraction of \textit{mass-matched sample} of green galaxies divided according to the transition mass as a function of the environment is given for the three green subclasses combined (C4--6) and separately (C4, 5, and 6). The $\delta$-bins corresponds to the quartiles of the $\delta$ distribution for the whole sample, and the fractions are normalised to the total number of galaxies in the same subclass, i.e. for each subclass the fractions in $\delta$ bins sum up to 100\%. The solid and dashed lines correspond to the weighted fit for low- ($9.5\lesssim\log ({\rm M_{star}/M}_{\odot})\lesssim10.6$) and high-mass ($10.6\lesssim\log ({\rm M_{star}/M}_{\odot})\lesssim11.5$) relation, respectively. Shaded stripes around lines display $1\sigma$ of the fit. The slope of the fit is given in the legend. The number of galaxies in low- and high-mass samples, respectively for each subclass are given in parenthesis.}
	\label{fig:green_fraction_mass}
\end{figure*} 

The fractions of green galaxies divided about the transition mass are shown in Fig.~\ref{fig:green_fraction_mass}. 
As there are no noticeable differences in the fraction-$\delta$ relations if we consider \textit{mass-matched samples} or not, we show plots only for the \textit{mass-matched samples}. 
In each panel, the green fraction as a function of the environment is given in two $\rm{M_{star}}$ bins for the three green subclasses combined (C4--6) and C4--6 separately.

Overall, green galaxies show a trend with $\rm{M_{star}}$ (transition mass) that may be correlated to downsizing. 
The low-mass green galaxies are defined here as the ones with $9.5\lesssim\log ({\rm M_{star}/M}_{\odot})\lesssim10.6$ (see Fig.~\ref{fig:green_massdistribution}). 
Our studies, however, do not extend to the dwarf regime ($\rm{log ( M_{star}/M}_{\odot})\lesssim9.5$), which are expected to be differently affected by environment~\citep[e.g.][]{Guo2017,moutard2018}  than more massive galaxies. 
They are not star-forming anymore (in fact they are intermediate) and their fraction tends to be flat with $\delta$. 
The lack of trend with $\delta$ suggests that low-mass green galaxies are not undergoing environmental quenching. 

Figure~\ref{fig:green_fraction_mass} shows that high-mass green galaxies have a strong preference towards high-$\delta$ regions, irrespective of subclass. 
This suggests that environmental quenching is primarily responsible for producing these high-mass green galaxies.
\cite{moutard2016a} showed that the high-mass green galaxies took longer to transit (pass through the green valley) than the low-mass green valley galaxies. 
This interpretation is supported by \cite{Haines2013} who found a large number of transition galaxies (with SFRs reduced by a factor 2-3) in clusters, whose quenching occurs slowly on 1--2 Gyr time-scales. 
The increasing trend of the fraction of massive green galaxies with increasing $\delta$ is also consistent with the hypothesis of merger-driven evolution of massive green valley galaxies discussed in~\cite{moutard2020}. 
The authors found that massive galaxies that are already in the process of quenching (i.e. green massive galaxies) are hosting heavily-obscured X-ray AGNs. This is consistent with a quenching scenario that involves major mergers. 

\section{Blue galaxies}\label{sec:starfroming}

The FEM classification yields  five blue star-forming subclasses, which catch the blue population at different parts along the galaxy main star-forming sequence (see Sec.~4.3 in \citetalias{Siudek2018a}). 
However, \citetalias{Siudek2018a} found that C11 contained many narrow-line AGNs, so we do not consider C11 in the remaining analysis, we discuss the environmental properties of this subclass in Siudek et al. in prep.

\begin{figure}
	\centerline{
	\includegraphics[width=0.49\textwidth]{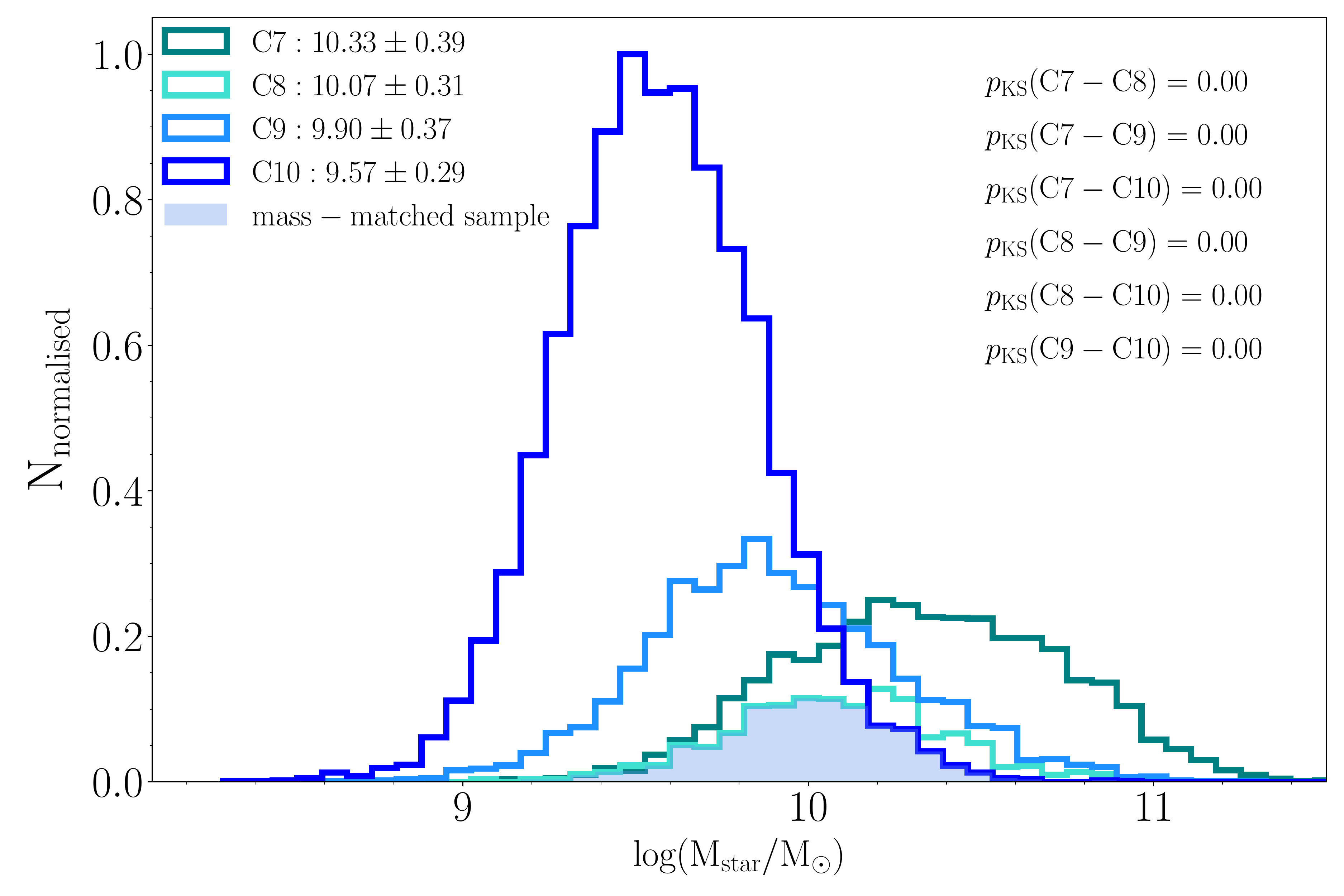}}
	\caption{$\rm{M_{star}}$ distributions of blue subclasses C7--10 normalised to the maximum value of the histograms. The mean and standard deviation is shown in the upper left. The KS probability is displayed in the upper right. The blue filled histogram indicates the $\rm{M_{star}}$ distribution for the \textit{mass-matched sample} and corresponds to the area below the distributions for all subclasses. The \textit{mass-matched sample} returns the same $\rm{M_{star}}$ distribution for all four C7--10 subclasses. }
	\label{fig:bluehisto}
\end{figure}

Figure~\ref{fig:bluehisto} shows the $\rm{M_{star}}$ distributions for C7--10. 
Following the approach introduced in Sec.~\ref{sec:redgalaxies}, we create the \textit{mass-matched sample} to check if the relation of the number or properties of blue subclasses depends on the different $\rm{M_{star}}$ distributions among C7--10. 
Blue subclasses are characterised by different mass distributions, as also confirmed by KS tests, and their median (see Tab.~\ref{table:eff}) and mean values (see the legend in Fig.~\ref{fig:bluehisto}). 
The \textit{mass-matched sample} of 3\,784 blue galaxies returns the same $\rm{M_{star}}$ distribution for C7--10 and allows us to directly compare the trends of the fraction-$\delta$ relations between the different subclasses. 

\subsection{The fraction-$\delta$ relation}

\begin{figure}
	\centerline{\includegraphics[width=0.49\textwidth]{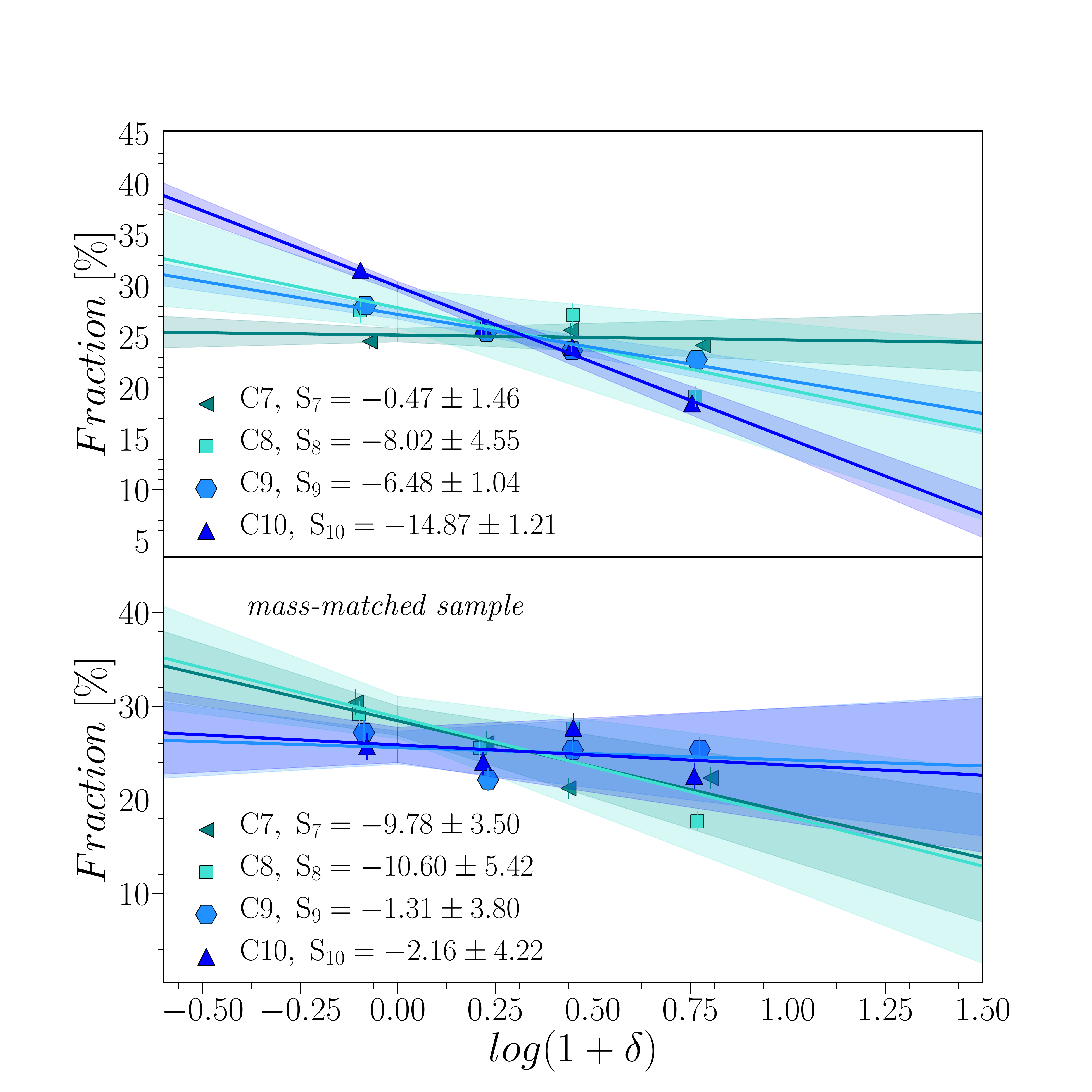}}
	\caption{Upper panel: The fraction of 3\,281, 1\,203, 3\,468, 9\,207 blue galaxies within C7, 8, 9, and 10, respectively, as a function of the $\delta$. The $\delta$-bins corresponds to the quartiles of the $\delta$ distribution for the whole sample and the fractions are normalised to the total number of galaxies in the same subclass, i.e. for each subclass, the fractions in $\delta$ bins sum up to 100\%. The solid line corresponds to the weighted fit. Shaded stripes around lines display $1\sigma$ of the fit. The slope of the fit is given in the legend. Bottom panel: As for the top panel, but for \textit{mass-matched sample}. The $\rm{M_{star}}$ distribution for each blue subclass is the same.}
	\label{fig:blue_fraction}
\end{figure}

The strong trend of the decreasing fraction of blue galaxies with a denser environment, visible in Fig.~\ref{fig:3class_fraction}, is driven mainly by the one (the most numerous) subclass - C10 (see Fig.~\ref{fig:blue_fraction}). 
The environmental dependence of the fraction of C7--9 is at least twice as weak as the one for C10, with C7 showing an almost flat trend. 
This may imply that around a half of the blue cloud population at $z\sim0.7$, gathered in C10 (see Tab.~\ref{table:eff}), follow the same, most likely mass-driven, secular SFH.
The remaining classes contain star-forming galaxies evolving under a combination of mass-dependent secular SFH and other physical processes, which leads to the lack of a strong correlation with the environment. 

Contrary to the behaviour of red and green subclasses, the blue one shows stronger dependence of the fraction-$\delta$ relation on their $\rm{M_{star}}$ distribution (see Fig.~\ref{fig:redfraction}, \ref{fig:green_fraction}, and \ref{fig:blue_fraction} for the trends found for red, green, and blue subclasses, respectively). 
The differences in stellar mass distributions of blue subclasses suggest that the low-mass ($\log ({\rm M_{star}/M}_{\odot})<10$) blue galaxies are the ones responsible for the strong decreasing trend of the fraction-$\delta$ relation, i.e. the dense-environment is penalising the survivors of low-mass galaxies.

\subsection{The size-$\delta$ relation}

\begin{figure}
	\centerline{\includegraphics[width=0.49\textwidth]{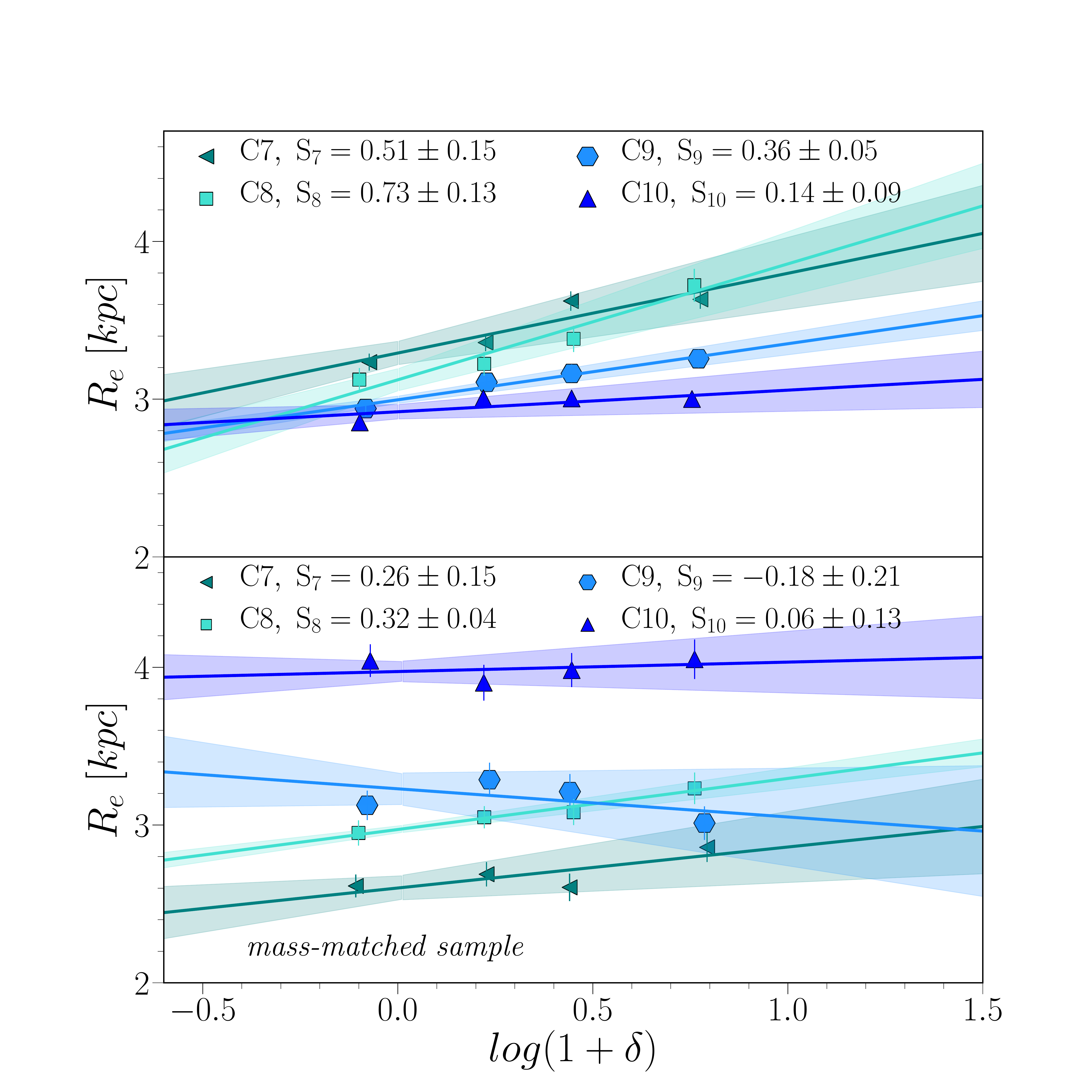}}
	\caption{Upper panel: Dependence of the $R_{e}$ on the $\delta$ for four blue galaxy subclasses.   The solid line corresponds to the weighted fit. Shaded stripes around lines display the $1\sigma$ uncertainty of the fit. 	 Bottom panel: As for the top panel, but for the \textit{mass-matched sample}. The $\rm{M_{star}}$ distribution for each blue subclass is the same.}
	\label{fig:blue_re}
\end{figure}

The size as a function of $\delta$ for blue subclasses C7--10 is shown in Fig.~\ref{fig:blue_re}.  
In the \textit{upper panel} in Fig.~\ref{fig:blue_re} we can see that, on average, $R_{e}$ of galaxies gathered in C10 is smaller and the trend of the increasing size with a denser environment is weaker than for galaxies within C7--9. 
$R_{e}$ of galaxies within C7--9 are increasing, on average, three times more than those of galaxies C10 when moving from low- to high-$\delta$ environments (i.e $R_{e}$ increases for $\sim15\%$, $\sim5\%$ from the lowest to the highest $\delta$ percentile for galaxies within C7--9, and C10, respectively). 

To verify if the difference in environmental trends of $R_e$ that we see between C7--9 and C10 are reflecting their different $\rm{M_{star}}$ distribution, we consider a \textit{mass-matched sample}(see the \textit{bottom panel} of Fig.~\ref{fig:blue_re}).  
The milder dependence of $R_{e}$ on the environment for \textit{mass-matched sample} for C10 than C7--9 may reflect that C10 gathers star-forming compact galaxies with the lowest $\rm{M_{star}}$ and even when found in high-$\delta$ environments, they remain small.  
\cite{Haines2007} showed that the effect of the environment on galaxy evolution increases when moving from high-mass to low-mass galaxies. 
Higher-mass star-forming galaxies are more resistant to external environmental processes than lower-mass
star-forming galaxies, and also evolve secularly (along the main star-forming sequence) earlier than dwarf galaxies. 

Clearly, we can see that the smaller $R_{e}$ found for the galaxies gathered in C10 (see \textit{upper panel} in Fig.~\ref{fig:blue_re}) were driven by lower-mass galaxies, as for the \textit{mass-matched sample} (\textit{lower panel}) C10 gathers the largest galaxies (with respect to remaining subclasses) with no signs of the preference of the size of galaxies  on the environment. 
This means that half of the blue population (i.e subclass C10) are compact low-mass galaxies preferentially residing in low-$\delta$ environments, but not showing different properties if found in denser regions. 

\subsection{Differences between blue subclasses}\label{sec:blue_difference}

Galaxies within C10 seem to be the star-forming galaxies that may follow the early, fast quenching channel described by~\citealt{moutard2016b,moutard2018} (see their Fig. 3a).  
Young, low-mass quiescent galaxies with bluer $r-K$ ($<$0.76) colours must have followed this fast quenching channel. 
This population of young quiescent galaxies is heavily affected/discarded by the double selection of VIPERS ($i<22.5$ and $z>0.5$), thus the FEM algorithm was not able to distinguish them as a separate subclass.  
However, their star-forming progenitors seem to be gathered in C10. 

Galaxies gathered in C10 are relatively low-mass (log$\rm{(M_{star}/M_{\odot})\sim9.56}$) and small ($R_{e}\sim2.96$~kpc).  
They have relatively high sSFR and preferentially reside in low-$\delta$ environments (see Tab.~\ref{table:eff}). 
This suggests that the most compact star-forming galaxies with high sSFRs are preferentially quenched in dense environments. 
This is in line with recent studies that galaxies with very high star formation activity are strongly depleted in dense environments because of the rapid environmental quenching (stellar feedback and the stripping of the extended halo), becoming post-starburst galaxies~\citep{Socolovsky2018, Socolovsky2019}. 

Galaxies within C7--9 may contain galaxies with different evolution mechanism than the most populated C10.  
The sSFRs are lower than for galaxies within C10, consistent with the hypothesis of fading as time passes. 
The sSFR was shown to depend on the environment, and on average, is lower in denser regions, which may be attributed to a lower fraction of star-forming galaxies in a denser environment~\citep[e.g.][]{peng2010,Darvish2016} and a lower star formation activity ($\sim 0.1-0.3$ dex) of star-forming galaxies in denser environments than the field~\citep[e.g.][]{Vulcani2010,Haines2013,Darvish2017}. 
At the same time, galaxies within C7--9 are larger than the ones from the blue population in C10 (see Tab.~\ref{table:eff}), showing larger dependence of their sizes on the environment. 
However, for the \textit{mass-matched sample} the effect is opposite: galaxies within C10 are larger than the ones in C7--9. 
The observed trend might imply that higher mass, larger galaxies can keep forming stars longer in hostile environments. 
\cite{Haines2017} showed that larger galaxies can continue to form stars to higher stellar masses than smaller galaxies.
That would suggest that the process of star formation is favoured or more resilient in more diffuse or less compact galaxies.

\section{Summary}\label{sec:summary}
We analyse the relations between the physical properties of galaxy (sub)classes and their $\delta$, based on a sample  
of 31\,631 VIPERS galaxies observed over redshift range $0.5<z<0.9$. 
In agreement with previous studies~\citep{capak2007, Tasca2009, Cucciati2017,PaulinoAfonso2019A&A...630A..57P} we  find, that the galaxy population-$\delta$ relation  is already at place at $z\sim0.9$. 
Based on the VIPERS sample, which has a statistical fidelity comparable to local surveys such as the SDSS but peaks at $z\sim0.7$, we confirm that in the whole analysed redshift range ($0.5<z<0.9$) the main trends seen in the local Universe \citep[e.g.][]{balogh2004, Baldry2006, Haines2007}, are already present  (see Fig.~\ref{fig:3class_fraction}). 
The fraction of red passive galaxies increases with $\delta$, while the fraction of blue star-forming galaxies decreases with increasing $\delta$.  
At the same time, red galaxies in high-$\delta$ environment are larger by $28\%$ than the ones in low-dense environments for the \textit{mass-matched sample}.
The galaxy population-$\delta$ relations at $0.7<z<0.9$ and $0.5<z<0.7$ are remarkably similar, suggesting that the processes correlated with environment played their  role to establish these trends at redshift higher than 1.

The more detailed classification allows us to have deeper insights into the influence of the environment on the properties of different populations of galaxies. 
We make use of the classification based on the unsupervised clustering method performed by~\citetalias{Siudek2018a} which subdivided the red, green, blue galaxy populations in VIPERS into 11 subclasses to test the hypothesis that these subclasses may be related to different evolutionary paths inside the three main galaxy populations. 
The most relevant findings are:
\begin{itemize}
    \item All three red subclasses (C1--C3) show the same preference to denser environments, however, their sizes differ and correlate with the local environment in different ways suggesting that dry merger activity shaped mostly galaxies within C3, while galaxies within C1 were quenched mostly as a result of internal processes.  
    \item High-mass ($\rm{log}(M_{star}/M_{\odot})$>10.6) green galaxies (C4--6) are driving the positive fraction-density relation found for the intermediate galaxies suggesting that environmental quenching is more important in the evolution of high-mass green galaxies 
    \item The majority of blue galaxies (C10) follow the decreasing trend of the fraction-density relation following genuine passive evolution through accretion of the surrounding gas.
\end{itemize}

\begin{acknowledgements}
The authors want to thank Bianca Garilli, Luigi Guzzo and the referee for useful and constructive comments. 
This work has been supported by the European Union's Horizon 2020 Research and Innovation programme under the Maria Sklodowska-Curie grant agreement (No. 754510), the Polish National Science Centre (UMO-2016/23/N/ST9/02963, and UMO-2018/30/M/ST9/00757), the Spanish Ministry of Science and Innovation through the Juan de la Cierva-formacion programme (FJC2018-038792-I), and by Polish Ministry of Science and Higher Education grant
DIR/WK/2018/12.
KM has been supported by the National Science Centre (UMO-2018/30/E/ST9/00082).
CPH acknowledges support from ANID through Fondecyt Regular 2021 project no. 1211909.

 \end{acknowledgements}
        
\bibliographystyle{aa}
\bibliography{vipers}

\appendix
\section{The $\delta$ distribution}\label{app:delta_distribution}
\begin{figure}[!h]
\hspace*{5cm}
	\centerline{\includegraphics[width=0.99\textwidth]{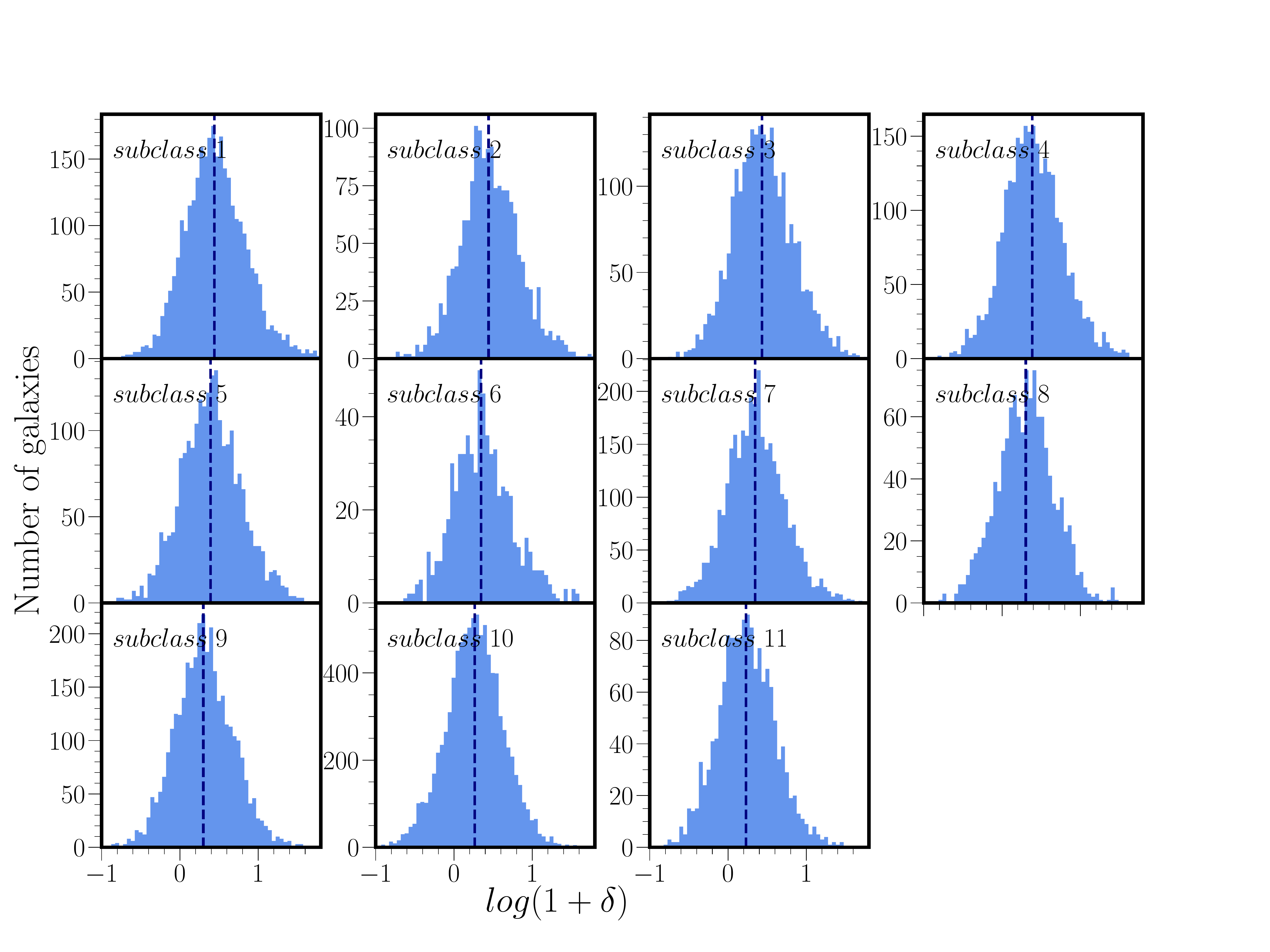}}	
	\caption{The distribution of over-densities for each VIPERS subclass. }
	\label{fig:delta_distribution}
\end{figure}

In Fig.~\ref{fig:delta_distribution} we show the distribution of $\rm{log}(1+\delta)$ for each VIPERS subclass. 

\end{document}